\documentclass[a4paper,11pt,leqno]{article}
\usepackage{amsmath}
\usepackage{amsfonts}
\usepackage{eucal}
\usepackage{amsthm}
\usepackage{amscd}

\def\b{\beta}

\def\ze{\mathbb{Z}}

\newtheorem{thm}{Theorem}
\newtheorem{lem}[thm]{Lemma}
\newtheorem{prop}[thm]{Proposition}

\theoremstyle{definition}

\theoremstyle{remark}
\newtheorem{rem}{Remark\!}

\begin{document}
\newpage\thispagestyle{empty}
{\topskip 2cm
\begin{center}
{\Large\bf Algebraic Topology of Spin Glasses\\}
\bigskip\bigskip
{\large Tohru Koma}\\
\bigskip
{\small \it Department of Physics, Gakushuin University, 
Mejiro, Toshima-ku, Tokyo 171-8588, JAPAN}\\
\smallskip
{\small\tt e-mail: tohru.koma@gakushuin.ac.jp}\\
\end{center}
\vfil
\noindent
We study topology of frustration in $d$-dimensional Ising spin glasses 
with $d\ge 2$ with nearest-neighbor interactions. We prove the following: 
For any given spin configuration, the domain walls on the unfrustration 
network are all transverse to a frustrated loop on the unfrustration network, 
where a domain wall is defined to be a connected element of the collection 
of all the $(d-1)$-cells which are dual to the bonds having an unfavorable 
energy, and the unfrustration network is the collection of all 
the unfrustrated plaquettes. These domain walls are topologically nontrivial 
because they are all related to the global frustration of a loop 
on the unfrustration network. Taking account of the thermal stability 
for the domain walls, we can explain the numerical results that 
three or higher dimensional systems exhibit a spin glass phase, 
whereas two-dimensional ones do not. Namely, in two dimensions, 
the thermal fluctuations of the topologically nontrivial domain walls destroy 
the order of the frozen spins on the unfrustration network, 
whereas they do not in three or higher dimensions. This may be interpreted 
as a global, topological effect of the frustrations. 
\par\noindent
\bigskip
\hfill
\vfil}\newpage

\section{Introduction}

Theoretical investigations of spin glasses have a long history, 
starting with the paper by Edwards and Anderson \cite{EA}. 
In particular, Toulouse \cite{Toulouse} emphasized the importance of 
frustration effect. Along the line, Bovier and Fr\"ohlich \cite{BF} studied the distribution 
of frustration, focusing on the geometrical aspect of frustration. 
However, such geometrical or topological approaches to spin glasses are still rare, 
and the nature of ordering at low temperatures still remains controversial, 
except for the success\footnote{See, for example, \cite{MPV}.} of the mean field theory.  

In this paper, we study topology of frustration 
in the Ising spin glasses with nearest-neighbor interactions 
on the $d$-dimensional hypercubic lattice $\ze^d$ with $d\ge 2$, 
in order to elucidate the nature of ordering in the spin glass phase at low temperatures. 
We show that algebraic topology is a useful mathematical language to describe the spin glasses. 
Actually, the frustration function which was introduced by Toulouse, 
is nothing but a one-dimensional cohomology class with $\ze_2$ coefficient 
as we will see in Section~\ref{Topunfrunet}. 

In order to describe our results, we introduce some basic terms. 
The more precise definitions are given in the following sections. 
Following Bovier and Fr\"ohlich \cite{BF}, we classify 
all the plaquettes into two classes, frustrated and unfrustrated plaquettes. 
The collection of the frustrated plaquettes is called the frustration network, and 
the latter collection is called the unfrustration network. 
For a given spin configuration, consider the collection of all the $(d-1)$-cells 
which are dual to the bonds having the unfavorable energy. 
The collection is made of the connected, $(d-1)$-dimensional hypersurfaces. 
We call a connected element of the hypersurfaces the domain wall. 

Now let us describe our results. When the system is restricted onto the unfrustration network, 
there exists a one-to-one correspondence 
between the homology class of the frustrated loops and the homology class of the domain walls. 
In more detail, each domain wall is transverse to the corresponding frustrated loop 
on the unfrustration network. These are topologically nontrivial domain walls 
because they are all related to the global frustration of a loop. 
For the whole lattice, the rest of the domain walls are outside of 
the unfrustration network. Thus there appear two types of the domain walls 
for a given spin configuration.

Based on these results, we can heuristically argue the existence of 
the spin glass phase at low temperatures. 
First we note that, in the same way as in \cite{BF}, one can show 
that there appears an infinite, connected cluster of unfrustrated plaquettes   
for a certain value of the concentration parameter for the positive couplings 
in dimensions $d\ge 2$. All the bonds of the unfrustration network except for 
the bonds which intersect a domain wall have a favorable energy. 
Clearly, in a ground state, the domain walls are determined to minimize 
their total size. Therefore the number of the bonds having an unfavorable energy 
on the unfrustration network becomes 
a small fraction of the whole bonds of the unfrustration network. 
{From} these observations, one notices that the situation on the unfrustration network 
is very similar to that in the standard ferromagnetic Ising model 
which is restricted onto the unfrustration network. 
The difference is the appearance of the two types of the domain walls. 

Consider first the effect of the domain walls which are outside of the unfrustration network. 
They are expected to affect the spin configurations on the unfrustration network 
as a boundary effect. As we will see in Section~\ref{topofru}, 
a ground state on a frustration network 
is highly degenerate. This suggests that the spins show disorder 
on the frustration network at finite temperatures. 
Therefore the boundary effect may behave as random fields \cite{vEMN,vENS} 
at the boundary of the unfrustration network. 
But the domain walls minimize their total size in a ground state. 
This implies that the domain walls show a tendency to confine themselves 
into a small neighborhood of the frustration network. 
As a result, the boundary effect is expected to be ignorably small.    

Next consider the effect of the domain walls which are transverse to a frustrated loop 
on the unfrustration network. These topologically nontrivial domain walls 
may be regarded as an analogue of Dobrushin domain wall \cite{Dobrushin}. 
However, the number of the domain walls increases as the size of the system increases. 
As is well known, a domain wall is unstable against the thermal fluctuation 
in two dimensions \cite{Gallavotti}, 
while it is stable in three and higher dimensions. 

{From} these observations, we can expect the following: 
In two dimensions, the thermal fluctuations of the topologically nontrivial domain walls 
destroy the order of the frozen spins on the unfrustration network. 
This explains the absence of the spin glass phase in two dimensions. 
Since these domain walls are related to the global frustration of a loop,  
destroying the order of the frozen spins may be interpreted as a topological effect. 

In three or higher dimensions, the thermal fluctuations of the topologically nontrivial 
domain walls cannot destroy the order of the frozen spins on the unfrustration network 
because of the stability of the domain walls. 
As a result, there appears long range order of the frozen spins 
on the unfrustration network at low temperatures. Namely the system exhibits 
a spin glass phase in three and higher dimensions at low temperatures.      

This paper is organized as follows. In the next section, we describe the models 
which we consider in the present paper, and establish some basic definitions 
which are related to frustration. In Section~\ref{Topunfrunet}, we study 
topology of generic unfrustration networks. As a result, we prove that 
the homology class of the frustrated loops is isomorphic to the homology 
class of the domain walls on the unfrustration network. 
This implies that the topology of the domain walls on the unfrustration network 
is uniquely determined by the frustrated loops on the unfrustration network, and that 
the topology is independent of spin configurations.  
In Section~\ref{topofru}, we study topology of certain frustration networks. 
Their ground states exhibit high degeneracy. 
In Section~\ref{linfru}, we study the relation between frustration networks themselves and 
frustrated loops on unfrustration networks. The resulting relation enable us to 
elucidate the homology of the domain walls for the whole lattice in Section~\ref{homoDW}. 
In Section~\ref{Perco}, we estimate the cluster sizes of the unfrustration network, 
and discuss the existence of long range order of the frozen spins 
on the unfrustration network at low temperatures. 

\section{Preliminaries}

We consider a short-range Ising spin glass with Hamiltonian \cite{EA},   
\begin{equation}
{\cal H}_\Lambda=-\sum_{\langle i,j\rangle} J_{ij}\sigma_i\sigma_j,
\label{ham}
\end{equation}
where $\sigma_i$ takes the values $\pm 1$, and the bond variables $\{J_{ij}\}$ form 
a family of independent, identically distributed random variables; the sum is over 
nearest neighbor pair $\langle i,j\rangle$ of the sites $i,j$ 
on\footnote{\label{cell}Although we can treat general lattices 
which can be identified with a CW-complex, 
we consider a finite subset $\Lambda$ of the $\ze^d$ lattice for simplicity. 
As general references for cell complexes or CW-complexes, see \cite{DFN,Spanier}.} 
the finite sublattice $\Lambda$ of the $d$-dimensional lattice $\ze^d$ with $d\ge 2$. 
We choose the distribution of $J=J_{ij}$ as \cite{BF} 
\[
d\rho(J)=[(1-x)g(-J)+xg(J)]dJ,
\]
where the concentration parameter $x$ satisfies $0\le x\le 1$, 
and $g$ is a nonnegative function 
with support on the positive reals, and satisfies normalization, 
\[
\int g(J)dJ=1.
\]
A fairly common choice for $g$ is the singular delta distribution, 
\begin{equation}
g(J)=\delta(J-J_0),\quad \mbox{with some}\ J_0>0.
\label{deltadistribution}
\end{equation}
The positive coupling $J$ is distributed according to a Bernoulli bond percolation 
process with density $x$. 

We denote by ${\hat J}_{ij}$ the sign of the coupling $J_{ij}$, 
i.e., ${\hat J}_{ij}=J_{ij}/|J_{ij}|$. Following Toulouse \cite{Toulouse}, 
we introduce ``frustration" which is defined by 
\[
\phi(\ell)=\prod_{\langle i,j\rangle\subset\ell}{\hat J}_{ij}
\]
for a loop (or a closed path) $\ell$ along the bonds of the lattice. 
(A loop $\ell$ is made of a collection of bonds.) 
When $\phi(\ell)$ takes the value $-1$, we say that the loop $\ell$ is frustrated. 
We denote by $p$ a plaquette 
(or elementary 2-cell\footnote{See Footnote~\ref{cell} for an $n$-cell.}) 
of the lattice. 
Clearly the boundary $\partial p$ of a plaquette $p$ is a loop 
which consists of the four bonds. 
When $\phi(\partial p)=-1$, we also say that the plaquette $p$ is frustrated. 
We call a collection of frustrated plaquettes a frustration network,  
and call a collection of unfrustrated plaquettes $p$ 
satisfying $\phi(\partial p)=+1$ an unfrustration network. 

Following Bovier and Fr\"ohlich \cite{BF}, we associate 
a frustration network with the collection of the $(d-2)$-cells in the dual lattice. 
Since a plaquette is a 2-cell, its dual is given by a $(d-2)$-cell 
in the dual lattice. For example, in two dimensions the dual to a plaquette 
is the center of gravity, and in three dimensions the dual is given by   
the bond in the dual lattice. {\em A frustration network is dual to 
a collection of $(d-2)$-cells which form $(d-2)$-dimensional complexes 
which are closed or whose boundaries end at the boundaries of 
the dual lattice $\Lambda^\ast$ to the lattice $\Lambda$ \cite{BF}.} 
To see this, consider a cube (or elementary 3-cell) $c$. 
Then one has 
\[
\prod_{p\subset c}\phi(\partial p)
=\prod_{\langle i,j\rangle\subset c}
\left({\hat J}_{ij}\right)^2=1.  
\]
This implies that an even number of the plaquettes must be frustrated in $c$. 
Thus no complex made of $(d-2)$-cells dual to frustrated 
plaquettes can end in any cube. 

\section{Topology of unfrustration networks}
\label{Topunfrunet}

In this section, we consider an unfrustration network ${\cal N}_+$. 
By definition, all the plaquettes $p$ of ${\cal N}_+$ satisfy 
$\phi(\partial p)=+1$ which is called the cocycle condition. 
In order to study the topology of ${\cal N}_+$, 
we rely on homology and cohomology theories.\footnote{As general references, 
see the books \cite{DFN,ES,Greenberg,Spanier}.}  

Let us denote by $b(\ell)$ the set of all the bonds in the loop $\ell$. 
Consider the symmetric difference of the two sets, $b(\ell_1)$ and $b(\ell_2)$, of 
the bonds, 
\[
\ell_1\ominus\ell_2:=\left(b(\ell_1)\backslash b(\ell_2)\right)\cup 
\left(b(\ell_2)\backslash b(\ell_1)\right), 
\]
for two loops, $\ell_1$ and $\ell_2$, in ${\cal N}_+$. 
If the symmetric difference $\ell_1\ominus\ell_2$ of the bonds consists of 
only four bonds of a single unfrustrated plaquette $p$ in ${\cal N}_+$, 
then one has 
\[
\phi(\ell_1)=\phi(\ell_2)
\]
by using the cocycle condition $\phi(\partial p)=+1$. 
For these two loops, we write $\ell_1\sim\ell_2$. 
Clearly this relation can be extended to a generic pair of two loops, 
$\ell$ and $\ell'$, in ${\cal N}_+$ when there exists a sequence of loops, 
$\ell=\ell_1,\ell_2,\ldots,\ell_n=\ell'$, satisfying the above 
condition $\ell_i\sim\ell_{i+1}$ for $i=1,2,\ldots,n-1$.     
For such two loops, we also write $\ell\sim\ell'$, and 
we say that the two loops, $\ell$ and $\ell'$, 
are homologous to each other. Besides we automatically obtain $\phi(\ell)=\phi(\ell')$ 
for $\ell\sim\ell'$ from the definitions. 
But we stress that the quantity $\phi(\ell)$ is not necessarily equal to $+1$. 
For any unfrustrated plaquette $p$, 
the boundary $\partial p$ is homologous to the empty set by definition. 
In this case, we write $\partial p\sim 0$, and more generally, 
we write $\ell\sim 0$ if the loop $\ell$ is homologous to the empty set. 
Clearly we have $\phi(\ell)=+1$ for $\ell\sim 0$. 

Consider a two-dimensional surface $s$ which is made of a collection of 
unfrustrated plaquettes. Then the boundaries $\partial s$ of $s$ become 
loops. From the above argument, if the symmetric difference $\ell_1\ominus\ell_2$ for 
two loops, $\ell_1$ and $\ell_2$, is equal to the boundaries $\partial s$ 
of a surface $s$, then one has $\ell_1\sim\ell_2$.  
We denote by $Z_1({\cal N}_+;\ze)$ the module (additive group) 
made of all the oriented loops (or cycles) with coefficients $\ze$, 
and denote by $B_1({\cal N}_+;\ze)$ the submodule made of the boundaries $\partial s$ 
for all the two-dimensional oriented surfaces $s$. Then 
the one-dimensional homology module $H_1({\cal N}_+;\ze)$ is defined 
to be the quotient module, 
\[
H_1({\cal N}_+;\ze):=Z_1({\cal N}_+;\ze)/B_1({\cal N}_+;\ze).
\]
This module is made of all the classes $[\ell]$ which is represented by 
a nontrivial loop $\ell\ /\hspace{-0.35cm}\sim 0$. 
The frustration $\phi$ yields the homomorphism,\footnote{In this case, a homomorphism is 
a map $f$ of an additive group $A$ into a multiplicative group $B$ 
such that $f(a+b)=f(a)f(b)$ for $a,b\in A$.}    
\[
\phi:H_1({\cal N}_+;\ze)\longrightarrow \ze_2, 
\]
by the relation $\phi(\ell_1)=\phi(\ell_2)$ for $\ell_1\sim\ell_2$.  

The following lemma plays a key role in the present paper. 

\begin{lem}
\label{lemma:nofrustration}
Fix the random variable $\{J_{ij}\}$ in the Hamiltonian ${\cal H}_\Lambda$ 
of (\ref{ham}) on a finite lattice $\Lambda$. 
Suppose that any two sites in $\Lambda$ are connected by a path of bonds 
in $\Lambda$, and suppose that any loop $\ell$ in $\Lambda$ satisfies 
$\phi(\ell)=+1$, i.e., no loop is frustrated in $\Lambda$. 
Then the Hamiltonian ${\cal H}_\Lambda$ of (\ref{ham}) has exactly two ground states.  
\end{lem}

\begin{rem}
There exist some similarities between the present spin glass 
and an electron gas in a periodic potential. 
The above lemma states that the vanishing of the frustration leads to 
the triviality of $\ze_2$ bundle for the spin glass.
The analogue in the electron gas is that the vanishing of the Chern number yields 
the triviality of U(1) bundle for the Bloch wavefunctions \cite{Panati}. 
This similarity comes from the gauge invariance of the two theories \cite{Toulouse}. 
\end{rem}

\begin{proof}
Let $i_0$ be a site in $\Lambda$. Then any site $j$ in $\Lambda$ can be 
connected with $i_0$ by a path $\gamma$ in $\Lambda$ from the assumption 
on $\Lambda$. Write $\gamma=\{i_0,i_1,\ldots,i_n=j\}$ 
by using the sequence $\{i_0,i_1,\ldots,i_n=j\}$ 
of the sites in $\Lambda$, where $\langle i_k,i_{k+1}\rangle$ are 
bonds of $\Lambda$ for $k=0,1,\ldots,n-1$. Fix the value 
of the Ising spin $\sigma_{i_0}$ at the site $i_0$. 
Then we can determine the value of the spin $\sigma_j$ at the site $j=i_n$ 
by using the relations ${\hat J}_{i_ki_{k+1}}\sigma_{i_k}\sigma_{i_{k+1}}=+1$ 
along the path $\gamma$. Clearly the value of the spin $\sigma_j$ is unique 
for the fixed value of $\sigma_{i_0}$ and the fixed path $\gamma$. 

In the same way, a different path $\gamma'$ in $\Lambda$ gives 
a value $\sigma_j'$ of the spin at the site $j$. We want to show $\sigma_j=\sigma_j'$. 
Namely the value of spin $\sigma_j$ is independent of paths. 
Consider the loop $\ell$ which consists of two path $\gamma$ and $-\gamma'$. 
Then we have $\phi(\ell)=+1$ from the assumption. This implies that,
in modulo 2 arithmetic,  
the number of the negative couplings $J_{st}$ in $\gamma$ is equal to that of 
the negative couplings $J_{s't'}$ in $\gamma'$. Therefore $\sigma_j=\sigma_j'$. 

As a result, the whole spin configuration on $\Lambda$ is uniquely 
determined by the fixed value of $\sigma_{i_0}$. 
Clearly the above condition, ${\hat J}_{i_ki_{k+1}}\sigma_{i_k}\sigma_{i_{k+1}}=+1$, implies 
that the resulting spin configuration is the ground state. 
Besides, the choice of the value $\sigma_{i_0}$ is exactly two. 
\end{proof}

In Lemma~\ref{lemma:nofrustration}, 
the assumption $\phi(\ell)=+1$ for any loop $\ell\subset\Lambda$ is too strong 
even for $\Lambda={\cal N}_+$ 
because a frustrated loop appears in a generic unfrustration network. 
Thus we must treat the subset $\{[\ell]\in H_1({\cal N}_+;\ze)\ |\ \phi(\ell)=-1\}$ of 
the homology module $H_1({\cal N}_+;\ze)$. For this purpose, it is convenient 
to rely on the cohomology theory on a generic lattice $\Lambda$ 
which is a collection of plaquettes. 
  
Consider bond variables $\tau=\{\tau_{ij}\}_{\langle i,j\rangle}$,  
where $\tau_{ij}$ takes the value $\pm 1$. 
We denote by $C_1(\Lambda;\ze)$ the module which is made of the oriented bonds (1-chains)
with coefficients $\ze$. An element $a\in C_1(\Lambda;\ze)$ is written 
\[
a=\sum_{\langle i,j\rangle}c_{\langle i,j\rangle}\langle i,j\rangle\quad
\mbox{with }\ c_{\langle i,j\rangle}\in\ze. 
\]
Then $\tau(a)$ is defined by 
\[
\tau(a)=\prod_{\langle i,j\rangle}\left(\tau_{ij}\right)^{c_{\langle i,j\rangle}}.
\] 
As usual, we define the coboundary operator $\partial^\ast$ by the adjoint of 
the boundary operator $\partial$ as 
\[
(\partial^\ast\tau)(s)=\tau(\partial s)
\] 
for a two-dimensional surface $s$. We denote by $Z^1(\Lambda;\ze_2)$ 
the set of all $\tau$ satisfying $\partial^\ast\tau=1$. 
By the linearity, the condition $\partial^\ast\tau=1$ is equivalent to 
the cocycle condition $\tau(\partial p)=1$ for any plaquette $p\subset\Lambda$. 
An element of $Z^1(\Lambda;\ze_2)$ is called a cocycle. 
We also denote by $B^1(\Lambda;\ze_2)$ the set of all $\tau$ such that 
$\tau$ has $\tau_{ij}=\epsilon_i\epsilon_j$ for all the bonds 
$\langle i,j\rangle$ with 
a site variable $\epsilon_i$ which takes the value $\pm 1$. 
Since the boundaries $\partial s$ of surfaces $s$ become the loops, 
one has $\partial^\ast\tau(s)=\tau(\partial s)=1$ 
for any $\tau\in B^1(\Lambda;\ze_2)$. 
This implies $B^1(\Lambda;\ze_2)\subset Z^1(\Lambda;\ze_2)$. 
If two cocycles $\tau,\tau'\in Z^1(\Lambda;\ze_2)$ satisfy 
$\tau_{ij}'=\tau_{ij}\epsilon_i\epsilon_j$ with site variables $\epsilon_i$, 
then one has $\tau'(\ell)=\tau(\ell)$ for any loop $\ell$. 
We say that such two cocycles, $\tau$ and $\tau'$, are cohomologous to 
each other, and write $\tau\sim\tau'$. 
(The two sets, $\tau,\tau'$, of the couplings are gauge equivalent to each other.) 
The one-dimensional cohomology module $H^1(\Lambda;\ze_2)$ is defined to be 
the quotient module of $Z^1(\Lambda;\ze_2)$ by the submodule $B^1(\Lambda;\ze_2)$. 

Now we return to the problem characterizing 
the collection of the frustrated loops, $\{[\ell]\in H_1({\cal N}_+;\ze)| \phi(\ell)=-1\}$.  
The universal coefficient 
theorem\footnote{For example, see Theorem~3 of Sec.~5.5 of \cite{Spanier}.} 
leads to the isomorphism,\footnote{A homomorphism is called an isomorphism if 
it is bijective.} 
\begin{equation}
{\rm Hom}\left(H_1({\cal N}_+;\ze),\ze_2\right)\cong H^1({\cal N}_+;\ze_2),
\label{frustratedloop}
\end{equation}
because the zero-dimensional homology module $H_0({\cal N}_+;\ze)$ with 
coefficients $\ze$ is free.\footnote{As is well known, 
$H_0({\cal N};\ze)\cong\ze\oplus\cdots\oplus\ze$ for any network ${\cal N}$.} 
Here ${\rm Hom}(A,B)$ stands for the set of the homomorphism $A\longrightarrow B$. 
A proof of (\ref{frustratedloop}) is given in Appendix~\ref{UCT}.
This statement is rephrased as follows: In general, a cohomology class of frustration 
$\phi$ given by a set of random bond variables $J_{ij}$ is equivalent to 
assigning a homology class of nontrivial loops (not homologous to zero) to frustration 
(minus sign) on the unfrustration network.   

Next we characterize the ground state of the Hamiltonian ${\cal H}_\Lambda$ 
on the unfrustration network $\Lambda={\cal N}_+$ by the topological invariants 
which are related to the frustration $\phi$.  
For this purpose, it is convenient to introduce the bond variable $\alpha_{ij}\in\{0,1\}$ 
which is defined by 
\[
\tau_{ij}=e^{i\pi\alpha_{ij}}
\]
for the bond variable $\tau_{ij}\in\{-1,1\}$. 
This yields the isomorphism $\kappa:\tau\longmapsto\alpha$ 
between the two cohomology modules. The one-dimensional cochain $\alpha=\kappa(\tau)$ 
is defined by 
\begin{equation}
\alpha=\sum_{\langle i,j\rangle}\alpha_{ij}\langle i,j\rangle^\ast, 
\label{1-cochain}
\end{equation}
where $\langle i,j\rangle^\ast(\langle m,n\rangle)=1$ 
if $\langle i,j\rangle=\langle m,n\rangle$, 
and $\langle i,j\rangle^\ast(\langle m,n\rangle)=0$ otherwise. 
By definition, one has 
\[
\tau(\ell)=\exp[i\pi \alpha(\ell)]\quad\mbox{with}\ \ 
\alpha(\ell)=\sum_{\langle i,j\rangle\subset\ell}\alpha_{ij}
\]
for a loop $\ell$. 

Consider the complex $\Sigma$ which consists of $(d-1)$-cells which are dual to 
the bonds $\langle i,j\rangle$ satisfying $\alpha_{ij}=1$ 
for a cocycle $\alpha\in Z^1({\cal N}_+;\ze_2)$. 
Here $Z^1({\cal N}_+;\ze_2)$ is the module of all the cochains $\alpha$ 
satisfying the cocycle condition $\alpha(\partial p)=0$ mod 2. 
By the cocycle condition, the complex $\Sigma$ cannot end 
in any plaquette $p\subset{\cal N}_+$. 
Namely the complex $\Sigma$ forms the $(d-1)$-dimensional hypersurfaces 
which are closed or whose boundaries are 
in the boundaries $\partial{\cal N}_+^\ast$ of the dual ${\cal N}_+^\ast$ of 
the unfrustration network ${\cal N}_+$. We write $\Sigma=\vartheta(\alpha)$   
for the homomorphism. The homomorphism $\vartheta$ yields the 
following isomorphism which is a special case of 
Poincar\'e-Lefschetz duality theorem:\footnote{For example, 
see Sec.~28 of \cite{Greenberg}, or Theorem~20 of Sec.~6.3 of \cite{Spanier}.}

\begin{prop}
\label{PLduality1}
The following isomorphism holds:     
\[
H^1({\cal N}_+;\ze_2)\cong H_{d-1}({\cal N}_+^\ast,\partial{\cal N}_+^\ast;\ze_2),
\] 
where the right-hand side is the set of the homology classes for 
the $(d-1)$-dimensional hypersurfaces which are closed or whose boundaries are 
in the boundaries $\partial{\cal N}_+^\ast$ of the dual ${\cal N}_+^\ast$ of 
the unfrustration network ${\cal N}_+$. 
\end{prop}

\begin{proof}
First we show that the map $\vartheta\circ\kappa$ is well defined, i.e.,  
if $\tau\sim\tau'$, then $\vartheta\circ\kappa(\tau)\sim\vartheta\circ\kappa(\tau')$. 
Let $\tau,\tau'\in Z^1({\cal N}_+;\ze_2)$ satisfying 
$\tau'\sim\tau$. 
Then $\tau_{ij}'=\tau_{ij}\epsilon_i\epsilon_j$ with the site variables 
$\epsilon_i\in\{+1,-1\}$, and  
$\vartheta\circ\kappa(\tau')=\vartheta\circ\kappa(\tau)
+\vartheta\circ\kappa(\epsilon)$, 
where $\epsilon=\{\epsilon_{ij}=\epsilon_i\epsilon_j\}$ by definition. 
Consider two regions, $V_+=\{i|\epsilon_i=+1\}$ and $V_-=\{i|\epsilon_i=-1\}$. 
Then the set of the bonds having $\epsilon_{ij}=-1$ can be identified with 
the interfaces $\Sigma$ between the two regions, $V_+$ and $V_-$ 
by using the duality of the lattices. On the other hands, 
these interfaces $\Sigma$ are nothing but the $(d-1)$-dimensional hypersurfaces 
$\vartheta\circ\kappa(\epsilon)$ by definition. 
Since the boundaries $\partial V_+$ of the region $V_+$ is written as 
$\partial V_+=\Sigma+\Sigma_0$ with a $(d-1)$-dimensional hypersurface 
$\Sigma_0\subset\partial{\cal N}_+^\ast$, 
one has $\vartheta\circ\kappa(\epsilon)\sim 0$.   
This implies $\vartheta\circ\kappa(\tau)\sim\vartheta\circ\kappa(\tau')$.

Let $\Sigma$ be a $(d-1)$-dimensional hypersurfaces which are closed 
or whose boundaries are in the boundaries $\partial{\cal N}_+^\ast$. 
If a plaquette $p\subset{\cal N}_+$ intersects $\Sigma$, 
an even number of the bonds of $p$ must intersect $\Sigma$. 
For such bonds $\langle i,j\rangle$, we choose the bond variables $\tau_{ij}=-1$.  
Then $\tau$ satisfies the cocycle condition, and $\vartheta\circ\kappa(\tau)=\Sigma$. 
Thus the map $\vartheta$ is surjective. 
 
Finally let us show that the corresponding map is injective. 
Let $\tau\in Z^1({\cal N}_+;\ze_2)$ satisfying $\vartheta\circ\kappa(\tau)\sim 0$. 
Then there exists a region $V$ such that 
$\partial V=\vartheta\circ\kappa(\tau)+\Sigma_0$ 
with $\Sigma_0\subset\partial{\cal N}_+^\ast$. 
We take the site variables $\epsilon_i$ that $\epsilon_i=-1$ for 
$i\in V$ and $\epsilon_i=+1$ for $i\notin V$. 

Let us show that the relation $\tau_{ij}\epsilon_i\epsilon_j=+1$ holds 
for all the bonds $\langle i,j\rangle$.
When both of the sites $i$ and $j$ are included in the region $V$, 
the bond $\langle i,j\rangle$ does not intersect the surface $\partial V$.  
Therefore one has $\tau_{ij}=+1$ and $\tau_{ij}\epsilon_i\epsilon_j=+1$.
Similarly, when neither $i$ nor $j$ is included in $V$, 
the bond $\langle i,j\rangle$ does not intersect $\partial V$, too.
Therefore, $\tau_{ij}=+1$ and $\tau_{ij}\epsilon_i\epsilon_j=+1$.
When the bond $\langle i,j\rangle$ intersects the surface $\vartheta\circ\kappa(\tau)$, 
one has $\tau_{ij}=-1$. In this case, one obtains $\epsilon_i\epsilon_j=-1$ 
because one of the two site, $i$ and $j$, is included in $V$ and the other is not. 
These two yield $\tau_{ij}\epsilon_i\epsilon_j=+1$.  
Thus the relation $\tau_{ij}\epsilon_i\epsilon_j=+1$ holds 
for all the bonds $\langle i,j\rangle$. This implies $\tau\sim 1$. 
\end{proof}

Combining this proposition with (\ref{frustratedloop}), one has the following 
isomorphism:
\begin{equation}
{\rm Hom}\left(H_1({\cal N}_+;\ze),\ze_2\right)\cong 
H_{d-1}({\cal N}_+^\ast,\partial{\cal N}_+^\ast;\ze_2)
\label{H1Hd-1}
\end{equation}
This gives the universal relation between the frustrated loops and 
the domain walls on the unfrustration network ${\cal N}_+$ as we will see below. 

Let $\beta$ be a cochain satisfying $\beta\sim\kappa(\phi)$ 
for the frustration $\phi$. Namely there exists $\psi$ such that 
$\beta=\kappa(\psi)$, and that $\psi\sim\phi$.  
By definition, the cochain $\beta$ is written 
in the above form (\ref{1-cochain}) as  
\[
\beta=\sum_{\langle i,j\rangle}\beta_{ij}\langle i,j\rangle^\ast 
\]
with the bond variables $\beta_{ij}\in\{0,1\}$. Write $b(\beta)$ for the set of 
the bonds $\langle i,j\rangle$ satisfying $\beta_{ij}=1$. 
Namely the bonds in $b(\beta)$ are dual to the $(d-1)$-cells 
of the hypersurfaces $\vartheta(\beta)$. We write $b(\beta)=\vartheta(\beta)^\ast$. 
Clearly we have $\beta(\ell)=0$ 
for any loop $\ell\subset{\cal N}_+\backslash\vartheta(\beta)^\ast$.  
This implies $\phi(\ell)=\psi(\ell)=\exp[i\pi\beta(\ell)]=1$ for any loop 
$\ell\subset{\cal N}_+\backslash\vartheta(\beta)^\ast$. 
Thus the frustration function $\phi$ is 
not frustrated on ${\cal N}_+\backslash\vartheta(\beta)^\ast$. 

Let $\{\sigma_i\in\{1,-1\}\}_{i\in{\cal N}_+}$ be a spin configuration 
on the sites of the unfrustration network ${\cal N}_+$.  
Let $\phi'$ be a frustration with the random variable 
${\hat J}_{ij}'={\hat J}_{ij}\sigma_i\sigma_j$ for the frustration $\phi$ 
with ${\hat J}_{ij}$. By definition, one has $\phi'\sim\phi$ and 
$\kappa(\phi')\sim\kappa(\phi)$. In addition, the corresponding $(d-1)$-dimensional 
hypersurfaces, $\vartheta\circ\kappa(\phi')$ and $\vartheta\circ\kappa(\phi)$, are 
homologous to each other, 
i.e., $\vartheta\circ\kappa(\phi')\sim\vartheta\circ\kappa(\phi)$ from 
Proposition~\ref{PLduality1}. 
Let $b_-(\phi')$ be the set of the bonds $\langle i,j\rangle$ satisfying 
${\hat J}_{ij}'={\hat J}_{ij}\sigma_i\sigma_j=-1$. 
By the definition of the map $\kappa$, one has $b_-(\phi')=b(\kappa(\phi'))$. 
Therefore the bonds in $b_-(\phi')$ are dual to the $(d-1)$-cells of 
the hypersurfaces $\vartheta\circ\kappa(\phi')$. If a loop $\ell$ satisfies 
$\phi(\ell)=-1$, then the number of the bonds of the set $b(\ell)\cap b_-(\phi')$ 
must be odd. This implies that the loop $\ell$ must intersect at least 
one of the $(d-1)$-cell of the hypersurfaces $\vartheta\circ\kappa(\phi')$. 
In order to prove this statement, 
let us assume that the number of the elements of $b(\ell)\cap b_-(\phi')$ 
is even. Then one has $(\kappa(\phi'))(\ell)=0$ mod~2. This implies 
\[
\phi(\ell)=\phi'(\ell)=\exp[i\pi(\kappa(\phi'))(\ell)]=1. 
\]
Thus this is a contradiction. 

This fact is rephrased as follows: 
All the domain walls in the collection $\vartheta\circ\kappa(\phi')$ are transverse 
to a frustrated loop $\ell$ on the unfrustration network ${\cal N}_+$. 
Here, we define a domain wall as follows: 
For a given spin configuration, consider the collection of the $(d-1)$-cells 
which are dual to the bonds $\langle i,j\rangle$ having ${\hat J}_{ij}\sigma_i\sigma_j=-1$ 
which yields the unfavorable energy for the bond. 
We call a connected element of the collection the domain wall. 
Clearly, the $(d-1)$-dimensional hypersurfaces $\vartheta\circ\kappa(\phi')$ are 
the set of the domain walls for the spin configuration $\{\sigma_i\}_{i\in{\cal N}_+}$. 

Take $\{\sigma_i\}_{i\in{\cal N}_+}$ to be a spin configuration of the ground state of 
the Hamiltonian ${\cal H}_{\tilde \Lambda}$ on 
the unfrustration network ${\tilde \Lambda}={\cal N}_+$.  
Combining the above observations, Lemma~\ref{lemma:nofrustration} and 
the isomorphism (\ref{frustratedloop}) or (\ref{H1Hd-1}), we obtain 
the following theorem which gives an algorithm to find the ground state: 

\begin{thm}
\label{thm:unfrusGS}
Fix the random variable $\{J_{ij}\}$ in the Hamiltonian (\ref{ham}). 
Let ${\cal N}_+$ be a unfrustration network 
which is made of a collection of unfrustrated plaquettes. 
Suppose that any two sites in ${\cal N}_+$ are connected by a path of bonds 
in ${\cal N}_+$. 
Then there exist $(d-1)$-dimensional connected 
hypersurfaces $\Sigma_1,\ldots,\Sigma_r\subset{\cal N}_+^\ast$ 
satisfying $\partial\Sigma_i\subset \partial{\cal N}_+^\ast$ for $i=1,\ldots,r$ such that 
each hypersurface $\Sigma_i$ is transverse to a homology class of frustrated loops 
in ${\cal N}_+$, 
and that the Hamiltonian ${\cal H}_\Lambda$ of (\ref{ham}) restricted to 
$\Lambda={\cal N}_+\backslash\{\Sigma_1^\ast,\ldots,\Sigma_r^\ast\}$ has 
exactly two ground states. 
Here $\Sigma_i^\ast$ is the set of the bonds which are dual to a $(d-1)$-cell of 
the hypersurfaces $\Sigma_i$. Further we can choose 
the hypersurfaces $\Sigma_1,\ldots,\Sigma_r$ so that 
the spin configurations of the two ground states become a ground-state spin configuration 
of the Hamiltonian ${\cal H}_{\tilde \Lambda}$ of (\ref{ham}) restricted to 
${\tilde \Lambda}={\cal N}_+$.  
\end{thm}

\begin{rem}
The hypersurfaces, $\Sigma_1,\ldots,\Sigma_r$, for the ground state 
never contains a connected element which is homologous to zero. 
In fact, if they contain such an element $\Sigma_i$, then 
flipping the spins inside the hypersurface $\Sigma_i$ lowers 
the energy of the ground state. The hypersurfaces, $\Sigma_1,\ldots,\Sigma_r$, are 
the domain walls for a ground-state spin configuration. Since the choice of the set of the 
hypersurfaces, $\Sigma_1,\ldots,\Sigma_r$, is not necessarily unique, 
the degeneracy of the ground state may be larger than two. 
In the infinite volume, we can expect that there appear many ground-state 
spin configurations with different domain walls because an infinitely large 
domain wall is stable against a local perturbation.  
\end{rem}

Although the following lemma is essentially due to Bovier and Fr\"ohlich \cite{BF}, 
it is most efficient when applied to a spin configuration of the ground state 
in Theorem~\ref{thm:unfrusGS}: 

\begin{lem}
\label{lem:int}
Fix the random variable $\{J_{ij}\}$ in the Hamiltonian (\ref{ham}). 
Let ${\cal N}_+$ be a unfrustration network 
which is made of a collection of unfrustrated plaquettes. 
Suppose that any two sites in ${\cal N}_+$ are connected by a path of bonds 
in ${\cal N}_+$. 
Let $S_0$ be the set of all the $(d-1)$-cells of 
the hypersurfaces, $\Sigma_1,\ldots,\Sigma_r$, 
for the ground state in Theorem~\ref{thm:unfrusGS}, 
and let $S_-$ be the set of all the $(d-1)$-cells 
which are dual to the bonds with the negative coupling $J_{ij}<0$.
(The collection of the cells in $S_-$ forms the hypersurfaces $\vartheta\circ\kappa(\phi)$ 
for the frustration $\phi$.) 
Then the symmetric difference, $(S_0\backslash S_-)\cup(S_-\backslash S_0)$, gives  
the interfaces between up and down spins in the ground-state spin configuration 
of the Hamiltonian on the unfrustration network ${\cal N}_+$. 
\end{lem}

\begin{proof}
We denote by $S_0^\ast$ the set of all the bonds which are dual to a cell in $S_0$, 
and $S_-^\ast$ the the set of all the bonds with the negative coupling $J_{ij}<0$. 

For $\langle i,j\rangle\notin S_0^\ast\cup S_-^\ast$,  
one has ${\hat J}_{ij}\sigma_i\sigma_j=+1$ and ${\hat J}_{ij}=+1$.  
These imply $\sigma_i=\sigma_j$. 
For $\langle i,j\rangle\in S_0^\ast\cap S_-^\ast$, 
one obtains ${\hat J}_{ij}\sigma_i\sigma_j=-1$ and ${\hat J}_{ij}=-1$. 
These yield $\sigma_i=\sigma_j$ again. Therefore no interface appears 
outside the symmetric difference, $(S_0\backslash S_-)\cup(S_-\backslash S_0)$. 

For $\langle i,j\rangle\in S_0^\ast\backslash S_-^\ast$, one obtains 
${\hat J}_{ij}\sigma_i\sigma_j=-1$ and ${\hat J}_{ij}=+1$. 
Therefore $\sigma_i=-\sigma_j$. The corresponding dual cell gives the element of 
the interfaces. For $\langle i,j\rangle\in S_-^\ast\backslash S_0^\ast$,
one has ${\hat J}_{ij}\sigma_i\sigma_j=+1$ and ${\hat J}_{ij}=-1$.
These yield $\sigma_i=-\sigma_j$ again. 
\end{proof}

For the concentration $x$ near $1/2$, we can expect that 
there appear many large interfaces between up and down spins 
in a ground-state spin configuration 
on a large unfrustration network.
If all the bonds which are dual to such a connected interface are 
removed from the network, the network is divided into two large parts.  
We will discuss this point again in Section~\ref{Perco} by relying 
on a percolation theory. 
As the concentration $x$ of the positive coupling increases beyond a critical 
value $x_c$, we can expect that there disappear large interfaces which divide 
a large unfrustration network into two large parts \cite{SBLB}. 
As a result, the spin glass phase changes to the ferromagnetic one \cite{ARS}.

\section{Topology of frustration networks}
\label{topofru}

In order to study frustration networks, we introduce a concept of 
an unfrustrated pair of frustrated plaquettes as follows. 
Let $p_1$ and $p_2$ be two frustrated plaquettes such that 
they share only a single bond. 
In this case, the union $p_1\cup p_2$ can be identified with 
a single unfrustrated plaquette whose boundary is made of the six bonds. 
Actually, one has 
\[ 
\phi(\partial(p_1\cup p_2))=\phi(\partial p_1)\phi(\partial p_2)=1
\]
for the frustration function $\phi$. We call such two plaquettes $p_1,p_2$ 
an unfrustrated pair of frustrated plaquettes. 
We write $e_{ij}=p_i\cap p_j$ for the common bond 
of frustrated plaquettes $p_i,p_j$ which form an unfrustrated pair.   

Consider a frustration network ${\cal N}_-$ which is a collection of 
unfrustrated pairs of frustrated plaquettes.
Let ${\cal B}_+({\cal N}_-)$ be the set of 
all the bonds $\langle i,j\rangle$ in ${\cal N}_-$ such that 
the bond $\langle i,j\rangle$ is not a common bond of two frustrated plaquettes 
which form an unfrustrated pair. Since the network ${\cal B}_+({\cal N}_-)$ is 
regarded as an unfrustration network which is made of the unfrustrated pairs, 
we can apply the argument in the preceding section to it.  
Let $\{\sigma_i\}_{i\in{\cal N}_-}$ be a spin configuration on 
the frustration network ${\cal N}_-$. 
Consider the random variables ${\hat J}_{ij}'={\hat J}_{ij}\sigma_i\sigma_j$ 
for the bonds $\langle i,j\rangle\in {\cal B}_+({\cal N}_-)$. 
Then the random variables ${\hat J}_{ij}'$ define 
the frustration function $\phi'$ which is cohomologous to the frustration $\phi$ 
on ${\cal B}_+({\cal N}_-)$. 
Let $b_-(\phi')$ be the set of the bonds $\langle i,j\rangle\in{\cal B}_+({\cal N}_-)$ satisfying 
${\hat J}_{ij}'={\hat J}_{ij}\sigma_i\sigma_j=-1$. 
The collection of the $(d-1)$-cells which are dual to the bonds $b_-(\phi')$ 
form the $(d-1)$-dimensional hypersurfaces. However, the $(d-1)$-cells of 
theses hypersurfaces are not necessarily connected in an unfrustrated 
pair of frustrated plaquettes. In order to recover the connectivity, 
we add some $(d-1)$-cells to the hypersurfaces as follows:  
If a frustrated plaquette in ${\cal N}_-$ has an odd number of 
the bonds $\langle i,j\rangle\in{\cal B}_+({\cal N}_-)$ 
satisfying ${\hat J}_{ij}'=-1$, then we take a unique $(d-1)$-cell for 
the corresponding unfrustrated pair so that the cell is dual to the common bond 
of the two plaquettes. The resulting set of the $(d-1)$-dimensional connected 
hypersurfaces, $\Sigma_1,\ldots,\Sigma_r$, 
satisfies $\partial\Sigma_i\subset\partial{\cal N}_-^\ast$.   
 
Let us choose a spin configuration $\{\sigma_i\}_{i\in{\cal N}_-}$ to be a ground 
state of the Hamiltonian on the frustration network ${\cal N}_-$.
In this case, we cannot expect that, in the ground state, 
there appears no domain wall (hypersurface) which is homologous to zero. 
In fact,  as we will show below, the existence of the common bonds $e_{ij}$ 
for the unfrustrated pairs can lower the energy of the bonds 
which are dual to a domain wall which is homologous to zero. 

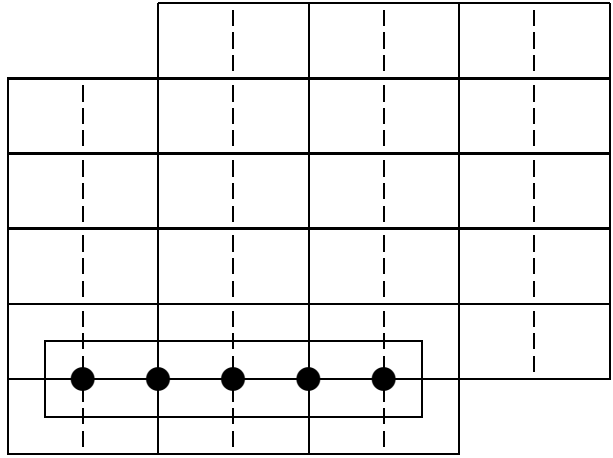
\begin{figure}
\setlength{\unitlength}{1mm}
\begin{center}
\begin{picture}(80,60)(0,0)
\put(0,0){\line(1,0){60}}
\put(0,10){\line(1,0){80}}
\put(0,20){\line(1,0){80}}
\put(0,30){\line(1,0){80}}
\put(0,40){\line(1,0){80}}
\put(0,50){\line(1,0){80}}
\put(20,60){\line(1,0){60}}
\put(0,0){\line(0,1){50}}
\put(20,0){\line(0,1){60}}
\put(40,0){\line(0,1){60}}
\put(60,0){\line(0,1){60}}
\put(80,10){\line(0,1){50}}
\put(10,1){\line(0,1){2}}
\put(10,4){\line(0,1){2}}
\put(10,7){\line(0,1){2}}
\put(10,11){\line(0,1){2}}
\put(10,14){\line(0,1){2}}
\put(10,17){\line(0,1){2}}
\put(10,21){\line(0,1){2}}
\put(10,24){\line(0,1){2}}
\put(10,27){\line(0,1){2}}
\put(10,31){\line(0,1){2}}
\put(10,34){\line(0,1){2}}
\put(10,37){\line(0,1){2}}
\put(10,41){\line(0,1){2}}
\put(10,44){\line(0,1){2}}
\put(10,47){\line(0,1){2}}
\put(30,1){\line(0,1){2}}
\put(30,4){\line(0,1){2}}
\put(30,7){\line(0,1){2}}
\put(30,11){\line(0,1){2}}
\put(30,14){\line(0,1){2}}
\put(30,17){\line(0,1){2}}
\put(30,21){\line(0,1){2}}
\put(30,24){\line(0,1){2}}
\put(30,27){\line(0,1){2}}
\put(30,31){\line(0,1){2}}
\put(30,34){\line(0,1){2}}
\put(30,37){\line(0,1){2}}
\put(30,41){\line(0,1){2}}
\put(30,44){\line(0,1){2}}
\put(30,47){\line(0,1){2}}
\put(30,51){\line(0,1){2}}
\put(30,54){\line(0,1){2}}
\put(30,57){\line(0,1){2}}
\put(50,1){\line(0,1){2}}
\put(50,4){\line(0,1){2}}
\put(50,7){\line(0,1){2}}
\put(50,11){\line(0,1){2}}
\put(50,14){\line(0,1){2}}
\put(50,17){\line(0,1){2}}
\put(50,21){\line(0,1){2}}
\put(50,24){\line(0,1){2}}
\put(50,27){\line(0,1){2}}
\put(50,31){\line(0,1){2}}
\put(50,34){\line(0,1){2}}
\put(50,37){\line(0,1){2}}
\put(50,41){\line(0,1){2}}
\put(50,44){\line(0,1){2}}
\put(50,47){\line(0,1){2}}
\put(50,51){\line(0,1){2}}
\put(50,54){\line(0,1){2}}
\put(50,57){\line(0,1){2}}
\put(70,11){\line(0,1){2}}
\put(70,14){\line(0,1){2}}
\put(70,17){\line(0,1){2}}
\put(70,21){\line(0,1){2}}
\put(70,24){\line(0,1){2}}
\put(70,27){\line(0,1){2}}
\put(70,31){\line(0,1){2}}
\put(70,34){\line(0,1){2}}
\put(70,37){\line(0,1){2}}
\put(70,41){\line(0,1){2}}
\put(70,44){\line(0,1){2}}
\put(70,47){\line(0,1){2}}
\put(70,51){\line(0,1){2}}
\put(70,54){\line(0,1){2}}
\put(70,57){\line(0,1){2}}
\put(10,10){\circle*{3}}\put(20,10){\circle*{3}}\put(30,10){\circle*{3}}
\put(40,10){\circle*{3}}\put(50,10){\circle*{3}}\put(5,5){\framebox(50,10){}}
\end{picture}
\end{center}
\caption{Two-dimensional lattice of unfrustrated pairs of frustrated plaquettes. 
Dotted lines indicate a common bond $e_{ij}$ of two frustrated plaquettes 
which form an unfrustrated pair.}
\label{2DLattice}
\end{figure}
\subsection{Frustration networks in two dimensions}

As a concrete example, consider a two-dimensional frustration network ${\cal N}_-$ 
in which the unfrustrated pairs form a subset of the $\ze^2$ lattice as 
in Figure~\ref{2DLattice}. For simplicity, we assume that 
the network ${\cal N}_-$ is simply connected, i.e., 
any loop in ${\cal N}_-$ is homologous to zero, and assume the delta distribution 
(\ref{deltadistribution}) for the random coupling $J_{ij}$, 
i.e., the bond variables $J_{ij}$ take the value $\pm J_0$. 
Following the above argument, one can find the spin configuration 
$\{\sigma_i\}_{i\in{\cal N}_-}$ satisfying $J_{ij}\sigma_i\sigma_j=J_0>0$ 
for all the bonds $\langle i,j\rangle\in{\cal B}_+({\cal N}_-)$.  
This implies that three bonds of each frustrated plaquette have 
the energy $J_{ij}\sigma_i\sigma_j=J_0>0$. Since the plaquette is 
frustrated, the rest has the energy $-J_0$.  For each unfrustrated pair, 
the common bond for the two plaquettes has the energy $-J_0$. From these 
observations, we obtain that the spin configuration $\{\sigma_i\}_{i\in{\cal N}_-}$ is 
one of the ground state of the Hamiltonian on ${\cal N}_-$. 
However, the ground states are highly degenerate. In fact, one can obtain 
another ground state by a local spin-flip. For example, 
flipping the five spins marked by the close circles in Figure~\ref{2DLattice} 
does not change the ground state energy. By this spin-flip, 
there appears the $(d-1)$-dimensional hypersurface which we expected above. 
In the present two-dimensional case,  
the hypersurface is the loop encircling the five spins, and is homologous to zero.   
Thus the hypersurface is not related to any frustrated loop. 

\subsection{Frustration networks in three dimensions}

The situation in three and higher dimensions is slightly different from 
that in two dimensions. 
To see this, let us consider the three-dimensional 
cubic lattice $\ze^3$. We assume that all the plaquettes are 
frustrated. We want to find a set of unfrustrated pairs of frustrated plaquettes 
so that each unfrustrated pair becomes a two-cell of the corresponding complex. 
We denote by ${\cal B}$ the set of all the bonds of the $\ze^3$ lattice. 
We define the subset ${\cal B}_-$ of the bonds as  
\[
{\cal B}_-:={\cal B}_{-,x}\cup{\cal B}_{-,y}\cup{\cal B}_{-,z},
\]
where the three sets of the bonds in the right-hand side are given by 
\[
{\cal B}_{-,x}:=\{\langle i,j\rangle\ |\ 
j=i+(1,0,0),\ i\in\ze\times 2\ze\times 2\ze\},
\]
\[
{\cal B}_{-,y}:=\{\langle i,j\rangle\ |\
j=i+(0,1,0),\ i\in 2\ze\times\ze\times 2\ze +(1,0,1)\},
\]
and 
\[
{\cal B}_{-,z}:=\{\langle i,j\rangle\ |\
j=i+(0,0,1),\ i\in 2\ze\times 2\ze\times \ze+(0,1,0)\}.
\]
Then we can take the set of unfrustrated pairs of frustrated plaquettes 
so that a common bond for frustrated plaquettes which form an unfrustrated pair 
is always an element of ${\cal B}_-$. Further we define a set ${\cal B}_+$ of bonds  
as ${\cal B}_+:={\cal B}\backslash{\cal B}_-$. This set ${\cal B}_+$ is 
the collection of the boundary bonds of the unfrustrated pairs. 
Each unfrustrated pair has six bonds in their boundary. 

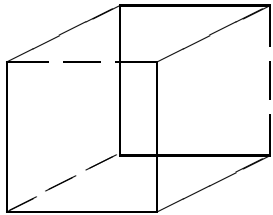
\begin{figure}
\setlength{\unitlength}{1mm}
\begin{center}
\begin{picture}(40,31)(0,0)
\put(5,0){\line(1,0){20}}
\put(5,0){\line(2,1){4}}\put(10.2,2.8){\line(2,1){4}}
\put(15.6,5.6){\line(2,1){4}}
\put(25,0){\line(2,1){15}}
\put(20,7.5){\line(1,0){20}}
\put(5,20){\line(1,0){5.5}}\put(12.5,20){\line(1,0){5}}\put(19.5,20){\line(1,0){5.5}}
\put(5,20){\line(2,1){15}}\put(25,20){\line(2,1){15}}
\put(20,27.5){\line(1,0){20}}
\put(5,0){\line(0,1){20}}\put(25,0){\line(0,1){20}}
\put(20,7.5){\line(0,1){20}}
\put(40,7.5){\line(0,1){5.5}}\put(40,15){\line(0,1){5}}\put(40,22){\line(0,1){5.5}}
\end{picture}
\end{center}
\caption{The boundary of a cube consists of three unfrustrated pairs of 
frustrated plaquettes. Dotted lines indicate a common bond $e_{ij}$ 
of two frustrated plaquettes which form an unfrustrated pair.}
\label{cube}
\end{figure}

As shown in Fig.~\ref{cube}, three bonds (dotted lines) of a cube 
are an element of ${\cal B}_-$, and the rest of the bonds are in ${\cal B}_+$.  
All the plaquettes of the cube are assumed to be frustrated. We can choose 
three unfrustrated pairs as two-cells of the cube. 
Namely the surface of the cube can be covered by 
the three unfrustrated pairs. From this observation, one can notice that 
any loop which is made of the boundary bonds of the unfrustrated pairs 
is homologous to zero in the cube.  
Figure~\ref{cube2} shows the case of two cubes whose two-cells 
are made of five unfrustrated pairs 
of frustrated plaquettes and one frustrated plaquette.
Since the two cubes have 11 frustrated plaquettes, their two-cells 
cannot be expressed in terms of unfrustrated pairs only. 
Figure~\ref{cube3} shows the case of four cubes. 
The two-cells are all made of unfrustrated pairs. 
We stress that the way of choosing unfrustrated pairs as two-cells is not unique. 
\begin{figure}
\setlength{\unitlength}{1mm}
\begin{center}
\begin{picture}(80,32)(-10,0)
\put(5,0){\line(1,0){20}}
\put(5,0){\line(2,1){4}}\put(10.2,2.8){\line(2,1){4}}
\put(15.6,5.6){\line(2,1){4}}
\put(25,0){\line(2,1){15}}
\put(20,7.5){\line(1,0){20}}
\put(5,20){\line(1,0){5.5}}\put(12.5,20){\line(1,0){5}}\put(19.5,20){\line(1,0){5.5}}
\put(5,20){\line(2,1){15}}\put(25,20){\line(2,1){15}}
\put(20,27.5){\line(1,0){20}}
\put(5,0){\line(0,1){20}}\put(25,0){\line(0,1){20}}
\put(20,7.5){\line(0,1){20}}
\put(40,7.5){\line(0,1){5.5}}\put(40,15){\line(0,1){5}}\put(40,22){\line(0,1){5.5}}
\put(25,0){\line(1,0){20}}
\put(45,0){\line(2,1){4}}
\put(50.2,2.8){\line(2,1){4}}
\put(55.6,5.6){\line(2,1){4}}
\put(40,7.5){\line(1,0){20}}
\put(25,20){\line(1,0){5.5}}\put(32.5,20){\line(1,0){5}}\put(39.5,20){\line(1,0){5.5}}
\put(25,20){\line(2,1){15}}
\put(45,20){\line(2,1){15}}
\put(40,27.5){\line(1,0){20}}
\put(25,0){\line(0,1){20}}\put(45,0){\line(0,1){20}}
\put(60,7.5){\line(0,1){20}}
\end{picture}
\end{center}
\caption{Two cubes whose two-cells are made of five unfrustrated pairs 
of frustrated plaquettes and one frustrated plaquette.}
\label{cube2}
\end{figure}
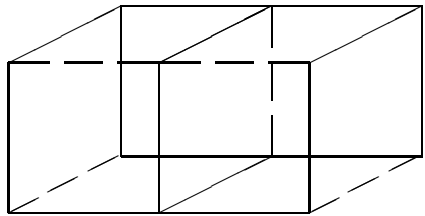
\begin{figure}
\setlength{\unitlength}{1mm}
\begin{center}
\begin{picture}(100,43)(-15,0)
\put(5,0){\line(1,0){20}}
\put(5,0){\line(2,1){4}}\put(10.2,2.8){\line(2,1){4}}
\put(15.6,5.6){\line(2,1){4}}
\put(25,0){\line(2,1){15}}
\put(20,7.5){\line(1,0){20}}
\put(5,20){\line(1,0){5.5}}\put(12.5,20){\line(1,0){5}}\put(19.5,20){\line(1,0){5.5}}
\put(5,20){\line(2,1){15}}\put(25,20){\line(2,1){15}}
\put(20,27.5){\line(1,0){20}}
\put(5,0){\line(0,1){20}}\put(25,0){\line(0,1){20}}
\put(20,7.5){\line(0,1){20}}
\put(40,7.5){\line(0,1){5.5}}\put(40,15){\line(0,1){5}}\put(40,22){\line(0,1){5.5}}
\put(25,0){\line(1,0){20}}
\put(45,0){\line(2,1){4}}
\put(50.2,2.8){\line(2,1){4}}
\put(55.6,5.6){\line(2,1){4}}
\put(40,7.5){\line(1,0){20}}
\put(25,20){\line(1,0){5.5}}\put(32.5,20){\line(1,0){5}}\put(39.5,20){\line(1,0){5.5}}
\put(25,20){\line(2,1){15}}
\put(45,20){\line(2,1){15}}
\put(40,27.5){\line(1,0){20}}
\put(25,0){\line(0,1){20}}\put(45,0){\line(0,1){20}}
\put(60,7.5){\line(0,1){20}}
\put(20,7.5){\line(2,1){4}}\put(25.2,10.3){\line(2,1){4}}
\put(30.6,13.1){\line(2,1){4}}
\put(40,7.5){\line(2,1){15}}
\put(35,35){\line(1,0){5.5}}\put(42.5,35){\line(1,0){5}}\put(49.5,35){\line(1,0){5.5}}
\put(5,20){\line(2,1){30}}\put(25,20){\line(2,1){30}}
\put(35,15){\line(1,0){40}}
\put(35,15){\line(0,1){20}}
\put(55,15){\line(0,1){20}}
\put(75,15){\line(0,1){20}}
\put(45,7.5){\line(0,1){5.5}}\put(40,15){\line(0,1){5}}\put(40,22){\line(0,1){5.5}}
\put(60,7.5){\line(2,1){4}}
\put(65.2,10.3){\line(2,1){4}}
\put(70.6,13.1){\line(2,1){4}}
\put(40,7.5){\line(1,0){20}}
\put(55,35){\line(1,0){5.5}}\put(62.5,35){\line(1,0){5}}\put(69.5,35){\line(1,0){5.5}}
\put(45,20){\line(2,1){30}}
\put(40,27.5){\line(1,0){20}}
\put(60,7.5){\line(0,1){20}}
\put(25,20){\circle*{2}}\put(40,27.5){\circle*{2}}\put(55,35){\circle*{2}}
\end{picture}
\end{center}
\caption{Four cubes whose two-cells are ten unfrustrated pairs of 
frustrated plaquettes. By reversing the spins marked by the close circles, 
another ground state can be obtained from the constructed one.}
\label{cube3}
\end{figure}
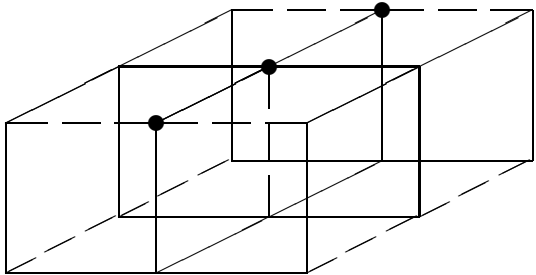

Now consider the $\ze^3$ lattice whose plaquettes are all frustrated. 
As we constructed above, all of the two-cells can be taken to   
be an unfrustrated pair of frustrated plaquettes. 
In order to avoid the difficulty coming from 
boundary or finite-size effects, we first consider the infinite-volume lattice 
although the situation is not realistic.   
Since the $\ze^3$ lattice with the bonds ${\cal B}_+$ is simply connected, 
there exists a spin configuration $\{\sigma_i\}_{i\in\ze^3}$ such that 
${\hat J}_{ij}\sigma_i\sigma_j=+1$ for all $\langle i,j\rangle\in{\cal B}_+$. 
When the random bond variables $J_{ij}$ take the values $\pm J_0$, 
this spin configuration is also a ground state of the Hamiltonian on $\ze^3$ 
with the bonds ${\cal B}={\cal B}_+\cup{\cal B}_-$. This is not a unique 
ground state because one can get another ground state by reversing 
the spins at the sites $\{i=(k,1,1)\ |\ k\in\ze\}$ on the line 
as in Fig.~\ref{cube3}. The sites are marked by the close circles in Fig.~\ref{cube3}. 
This implies that the ground state is highly degenerate. 
But, in contrast to two dimensions, the ground states show stability for 
local perturbations as: 

\begin{prop}
\label{prop:3Dstability}
Assume that the random bond variables $J_{ij}$ take the values $\pm J_0$, 
and assume that all of the plaquettes on $\ze^3$ are frustrated. 
If a spin configuration satisfies that the sum of the bond energies 
for each cube is minimized, then the spin configuration is a ground state 
of the Hamiltonian on $\ze^3$. Further the ground state is stable against any 
local spin-flip, i.e., any local spin-flip costs a finite energy for 
the ground state.  
\end{prop}

\begin{proof}
Write $\{\sigma_i\}_{i\in\ze^3}$ for the spin configuration. For each cube, 
three bonds must have the energy $J_{ij}\sigma_i\sigma_j=-J_0$ 
because all the plaquettes are frustrated. 
Let $V$ be a finite region in which the spins are flipped. The boundary $\partial V$ 
of $V$ is two-dimensional surface which encloses the spins on $V$. 
One can find a cube $c$ such that only a single site of $c$ is in $V$ and 
that the remaining seven sites of $c$ are outside $V$. 
Clearly only three bonds of $c$ intersect the boundary $\partial V$ of $V$. 
By the spin-flip, the three bonds change the energy. As a result, 
the cube $c$ must have at least four bonds having the energy $-J_0$.  
This increases the energy of the ground state.   
\end{proof}

For a finite subset of the frustration network 
in Proposition~\ref{prop:3Dstability}, the stability of a ground-state 
spin configuration does not hold because of the existence of the boundaries. 
As to thermal fluctuations, we can expect the global spin-flip symmetry 
to remain unbroken at finite temperatures for the spin system 
having the frustration network on $\ze^3$ 
because we can flip spins on a large size cluster 
with a small energy cost as we have seen in the above argument. 
Thus we can expect that finite-size effects and thermal fluctuations  
make spin configurations on frustration networks disorder. 

\section{Links of frustrations}
\label{linfru}

In this section, we study the relation between frustration networks themselves and 
frustrated loops in unfrustration networks. 

As we showed in Theorem~\ref{thm:unfrusGS}, all the domain walls in 
a ground state of the Hamiltonian on an unfrustration network ${\cal N}_+$ 
are transverse to a frustrated loop in ${\cal N}_+$. 
Each domain wall is a connected element of the collection of the $(d-1)$-dimensional hypersurfaces 
whose boundaries are included in the boundaries $\partial {\cal N}_+^\ast$ of 
the dual lattice ${\cal N}_+^\ast$ of ${\cal N}_+$. 
Let $\Sigma$ be such a domain wall, and let $\ell$ be the frustrated loop 
which is transverse to the domain wall $\Sigma$. 

Consider first the case of $\partial\Sigma\ne\emptyset$. 
The boundary $\partial\Sigma$ is the $(d-2)$-dimensional closed complex 
in $\partial {\cal N}_+^\ast$. Since $\partial\Sigma$ is outside ${\cal N}_+$, 
each $(d-2)$-cell of $\partial\Sigma$ is dual to a plaquette which is 
inside a frustrated network ${\cal N}_-$ or outside the lattice $\Lambda$ 
on which the Hamiltonian is defined. 
Namely we have $\partial\Sigma\subset{\cal N}_-^\ast\cup\partial\Lambda^\ast$. 
Since the frustrated loop $\ell$ is transverse to the domain wall $\Sigma$, one has 
the linking number, 
\[
{\rm Link}(\ell,\partial\Sigma)=1\ \ \mbox{mod}\ 2,
\]
when $\partial\Sigma\ne\emptyset$. 

If a two-dimensional surface $s$ satisfies $\partial s=\ell$, 
then the $(d-2)$-complex $\Gamma=\partial\Sigma$ is transverse to 
the two-dimensional surface $s$. However, it is not clear whether or not 
there exists a two-dimensional surface $s$ satisfying $\partial s=\ell$. 
In fact, one can easily construct a ground-state spin configuration 
with a single domain wall on the two-dimensional torus whose plaquettes are all unfrustrated, 
and the boundary $\partial\Sigma$ of the domain wall is vanishing. 
In this example, there exists no surface whose boundary gives the frustrated loop. 
The existence of the non-trivial frustrated loop is a consequence of 
the topology of the torus. Instead of the torus, if we consider a rectangular 
box with a free boundary, we can expect the existence of a surface whose 
boundary is a frustrated loop.

Since the frustrated loop $\ell$ is defined on an unfrustration network ${\cal N}_+$, 
we use the relative homology theory\footnote{See, for example, Section~13 of the book 
\cite{Greenberg}.} on the union ${\cal N}_-\cup{\cal N}_+$ of the two networks. 
We denote by $Z_2({\cal N}_-\cup{\cal N}_+,{\cal N}_+;\ze)$ 
the module which is made of all the two-dimensional surfaces $s$ 
satisfying $\partial s\subset{\cal N}_+$.  
We also denote by $B_2({\cal N}_-\cup{\cal N}_+,{\cal N}_+;\ze)$
the module which is made of all the two-dimensional surfaces $s$ 
which is written $s+s_0=\partial v$, where the two-dimensional surface $s_0$ 
is included in ${\cal N}_+$, and $v$ is a three-dimensional complex.  
Clearly the module $B_2({\cal N}_-\cup{\cal N}_+,{\cal N}_+;\ze)$ is a subset of 
$Z_2({\cal N}_-\cup{\cal N}_+,{\cal N}_+;\ze)$. 
The two-dimensional homology module $H_2({\cal N}_-\cup{\cal N}_+,{\cal N}_+;\ze)$ 
is defined to be the quotient module of $Z_2({\cal N}_-\cup{\cal N}_+,{\cal N}_+;\ze)$ 
by the submodule $B_2({\cal N}_-\cup{\cal N}_+,{\cal N}_+;\ze)$. 

As is well known, the following lemma holds: 

\begin{lem}
\label{lem:homoexact}
The following  sequence is exact: 
\begin{multline*}
H_2({\cal N}_-\cup{\cal N}_+;\ze)\mathop{\longrightarrow}^{j_\ast}
H_2({\cal N}_-\cup{\cal N}_+,{\cal N}_+;\ze)\mathop{\longrightarrow}^\partial
H_1({\cal N}_+;\ze)\\ \mathop{\longrightarrow}^{i_\ast}
H_1({\cal N}_-\cup{\cal N}_+;\ze)
\end{multline*}
Namely the image of the inclusion $j_\ast$ is equal to the kernel 
of the boundary map $\partial$, and the image of $\partial$ 
is equal to the kernel of the inclusion $i_\ast$. 
\end{lem}

\begin{proof}
(i) First let us show ${\rm Im}\ j_\ast={\rm Ker}\ \partial$. 
Clearly one has ${\rm Im}\ j_\ast\subset{\rm Ker}\ \partial$. 
In order to show ${\rm Im}\ j_\ast\supset{\rm Ker}\ \partial$, 
let $s$ be the two-dimensional surfaces $s$ satisfying 
$\partial s\sim 0$ in ${\cal N}_+$.  One has $\partial s=\partial s_0$ 
with a two-dimensional surface $s_0\subset{\cal N}_+$. 
Immediately $\partial(s-s_0)=0$. This implies that the homology class $[s-s_0]$ is 
an element of $H_2({\cal N}_-\cup{\cal N}_+;\ze)$. 
Besides, $s-s_0\sim s$ in $H_2({\cal N}_-\cup{\cal N}_+,{\cal N}_+;\ze)$. 
\smallskip

\noindent
(ii)  Next we show ${\rm Im}\ \partial={\rm Ker}\ i_\ast$. One can easily obtain 
${\rm Im}\ \partial\subset{\rm Ker}\ i_\ast$. Let $\ell$ be a loop in ${\cal N}_+$ 
that $\ell\sim 0$ in ${\cal N}_-\cup{\cal N}_+$. Then there exists a surface $s$ in 
${\cal N}_-\cup{\cal N}_+$ such that $\ell=\partial s$. Thus $\ell$ is within the image of 
$\partial$.  
\end{proof}

\noindent
{From} this lemma, when ${\cal N}_-\cup{\cal N}_+$ is simply connected, 
there exists a two-dimensional surface $s$ in ${\cal N}_-\cup{\cal N}_+$ 
such that $\partial s=\ell$ for any given frustrated loop $\ell\subset{\cal N}_+$. 
The surface $s$ must contain an odd number of frustrated plaquettes 
because the loop $\ell=\partial s$ is frustrated. 
Correspondingly there are the odd number of $(d-2)$-complexes which pierce 
the surface $s$. One of them, the boundary $\partial\Sigma$, pierces  
the two-dimensional surface $s$, too, because 
the loop $\ell$ is transverse to the domain wall $\Sigma$. 
Since the frustration of the loop $\ell$ is not 
affected by an even number of the frustrated plaquettes, 
we can take $\Gamma=\partial\Sigma$ to be the representative of 
the frustration which affects the unfrustration network ${\cal N}_+$. 

These observations suggest 
the existence of a relation between a frustration function of the plaquettes 
and the $(d-2)$-complex $\Gamma=\partial\Sigma$ for a domain wall $\Sigma$ 
on the network ${\cal N}_-\cup{\cal N}_+$. 
In order to explore the relation, we rely on 
the relative cohomology theory.\footnote{See, for example, the book \cite{Greenberg}.} 
Since the frustrated loop $\ell$ is defined on an unfrustration network, 
we consider the union ${\cal N}_-\cup{\cal N}_+$ of the two networks, ${\cal N}_-$ and ${\cal N}_+$, 
where ${\cal N}_+$ is the collection of 
all the unfrustration networks which touch ${\cal N}_-$. 
We do not require the connectivity of ${\cal N}_-$ and of ${\cal N}_+$. 
Let us introduce plaquette variables $\eta=\{\eta_p\}_p$ 
for all the plaquettes, where $\eta_p$ takes the values $\pm 1$.
Then the two-cochain $\eta$ is defined by  
\[
\eta(s)=\prod_{p\subset s}\eta_p
\]
for a two-complex $s$. Since the bond variables ${\hat J}_{ij}$ defines 
the plaquette variables $\eta_p$ by 
\[
\eta_p=\prod_{\langle i,j\rangle\subset p}{\hat J}_{ij},
\]
one can define the two-cochain $\Phi$ for the bond variables ${\hat J}_{ij}$ by 
\[
\Phi(s)=\prod_{p\subset s}\prod_{\langle i,j\rangle\subset p}{\hat J}_{ij}
=\prod_{\langle i,j\rangle\subset\partial s}{\hat J}_{ij}
\]
for a two-complex $s$. Since $\partial s$ is a collection of loops, 
we have $\Phi(s)=\phi(\partial s)$ with the frustration $\phi$. 
This yields the homomorphism $\partial^\ast:\phi\longmapsto\Phi$. 
(See Lemma~\ref{exactcoho} below.)
Clearly the two-cochain $\Phi$ satisfies the condition $\Phi(s)=1$ for $s\subset{\cal N}_+$. 

Let us go back to the general setting for two-cochains. 
We denote by $Z^2({\cal N}_-\cup{\cal N}_+;\ze_2)$ 
the two-dimensional cohomology module which is made of all the two-cochains $\eta$ 
which satisfy the cocycle condition $\eta(\partial c)=1$ for 
any cube $c$. We also denote by $B^2({\cal N}_-\cup{\cal N}_+;\ze_2)$ 
the module which is made of all the two-cochains $\eta$ which are given by 
\begin{equation}
\label{defB2}
\eta_p=\prod_{\langle i,j\rangle\subset p}\epsilon_{ij}
\end{equation}
with bond variables $\epsilon_{ij}$ which take the values $\pm 1$. 
The two-dimensional cohomology module 
$H^2({\cal N}_-\cup{\cal N}_+;\ze_2)$ is defined to be the quotient module 
$Z^2({\cal N}_-\cup{\cal N}_+;\ze_2)$ by the submodule 
$B^2({\cal N}_-\cup{\cal N}_+;\ze_2)$. 
Further we denote by $Z^2({\cal N}_-\cup{\cal N}_+,{\cal N}_+;\ze_2)$ 
the module which is made of all the two-cocycles $\eta\in Z^2({\cal N}_-\cup{\cal N}_+;\ze_2)$ 
satisfying the condition $\eta(p)=1$ for $p\subset{\cal N}_+$, 
and $B^2({\cal N}_-\cup{\cal N}_+,{\cal N}_+;\ze_2)$ 
the module which is made of all the two-cocycles $\eta\in B^2({\cal N}_-\cup{\cal N}_+;\ze_2)$ 
with the bond variables $\{\epsilon_{ij}\}$ of (\ref{defB2}) 
satisfying the condition $\epsilon_{ij}=1$ for $\langle i,j\rangle\subset{\cal N}_+$.
{From} the definitions, one has 
\[
B^2({\cal N}_-\cup{\cal N}_+,{\cal N}_+;\ze_2)\subset
Z^2({\cal N}_-\cup{\cal N}_+,{\cal N}_+;\ze_2).
\] 
Similarly, the two-dimensional cohomology module 
$H^2({\cal N}_-\cup{\cal N}_+,{\cal N}_+;\ze_2)$ is defined to be the quotient module as 
\[
H^2({\cal N}_-\cup{\cal N}_+,{\cal N}_+;\ze_2):=
Z^2({\cal N}_-\cup{\cal N}_+,{\cal N}_+;\ze_2)/B^2({\cal N}_-\cup{\cal N}_+,{\cal N}_+;\ze_2). 
\]
If $\eta,\eta'\in Z^2({\cal N}_-\cup{\cal N}_+,{\cal N}_+;\ze_2)$ satisfy  
the relation $\eta_p'=\eta_p\epsilon_p$ with 
$\epsilon=\{\epsilon_p\}_p\in B^2({\cal N}_-\cup{\cal N}_+,{\cal N}_+;\ze_2)$ 
for any plaquette $p$, 
then one has $\eta'(s)=\eta(s)$ for any two-dimensional surface $s$ 
satisfying $\partial s\subset{\cal N}_+$. 
Let $s,s'$ be two-dimensional surfaces which satisfy 
$s-s'\in B_2({\cal N}_-\cup{\cal N}_+,{\cal N}_+;\ze)$. 
Namely there exists 
a two-dimensional surface $s_0\subset{\cal N}_+$ and a three-dimensional complex $v$ 
such that $s-s'+s_0=\partial v$. Then one has $\eta(s)=\eta(s')$ 
for any $\eta\in Z^2({\cal N}_-\cup{\cal N}_+,{\cal N}_+;\ze_2)$.
Thus a cohomology class $[\eta]\in H^2({\cal N}_-\cup{\cal N}_+,{\cal N}_+;\ze_2)$ 
defines the homomorphism: 
\[
[\eta]:H_2({\cal N}_-\cup{\cal N}_+,{\cal N}_+;\ze)\longrightarrow \ze_2
\]

The cohomology version of Lemma~\ref{lem:homoexact} is: 

\begin{lem}
\label{exactcoho}
The following sequence is exact: 
\begin{multline*}
H^1({\cal N}_-\cup{\cal N}_+;\ze_2)\mathop{\longrightarrow}^{i^\ast}
H^1({\cal N}_+;\ze_2)\mathop{\longrightarrow}^{\partial^\ast}
H^2({\cal N}_-\cup{\cal N}_+,{\cal N}_+;\ze_2)\\ \mathop{\longrightarrow}^{j^\ast}
H^2({\cal N}_-\cup{\cal N}_+;\ze_2)
\end{multline*}
Here $i^\ast$ and $j^\ast$ are the inclusion maps. 
\end{lem}

\begin{proof}
(i) From the definitions, one has ${\rm Im}\ i^\ast\subset{\rm Ker}\ \partial^\ast$. 
In order to show ${\rm Im}\ i^\ast\supset{\rm Ker}\ \partial^\ast$, 
let $\tau\in Z^1({\cal N}_+:\ze_2)$ satisfy $\partial^\ast\tau\sim 1$. 
Then one has 
\[
(\partial^\ast\tau)(p)=\prod_{\langle i,j\rangle\subset p}\tau_{ij}
=\prod_{\langle i,j\rangle\subset p}\epsilon_{ij}
\]
for any plaquette $p$, where the bond variables $\epsilon_{ij}$ satisfy 
the condition $\epsilon_{ij}=1$ for $\langle i,j\rangle\subset{\cal N}_+$.
Set $\tau_{ij}'=\tau_{ij}\epsilon_{ij}$. Then the corresponding cocycle 
$\tau'$ is an element of $Z^1({\cal N}_-\cup{\cal N}_+;\ze_2)$, and 
$\tau_{ij}'=\tau_{ij}$ for $\langle i,j\rangle\subset{\cal N}_+$. 
Further one has $(\partial^\ast\tau')(p)=1$ for any plaquette $p$. 
\smallskip

\noindent
(ii) Clearly one has ${\rm Im}\ \partial^\ast\subset{\rm Ker}\ j^\ast$. 
Let us show ${\rm Im}\ \partial^\ast\supset{\rm Ker}\ j^\ast$. 
Let $\eta\in Z^2({\cal N}_-\cup{\cal N}_+,{\cal N}_+;\ze_2)$ satisfy 
$(j^\ast\eta)\sim 1$. Then there exist bond variables $\epsilon_{ij}$ which satisfy    
\[
(j^\ast\eta)_p=\prod_{\langle i,j\rangle\subset p}\epsilon_{ij}.
\]
{From} $\eta\in Z^2({\cal N}_-\cup{\cal N}_+,{\cal N}_+;\ze_2)$, one has 
$(j^\ast\eta)_p=\eta_p=1$ for $p\subset{\cal N}_+$. Combining these, one obtains 
\[
\prod_{\langle i,j\rangle\subset p}\epsilon_{ij}=1 \quad\mbox{for }\ p\subset{\cal N}_+.
\]
This implies that $\epsilon=\{\epsilon_{ij}\}\in Z^1({\cal N}_+;\ze_2)$ and that 
$\eta=\partial^\ast\epsilon$.  
\end{proof}

The following proposition is a special case of 
Poincar\'e-Lefschetz duality, too. 
This is nothing but the relation which we have explored, 
i.e., the relation between the frustration function of plaquettes 
and $(d-2)$-complexes. 

\begin{prop}
\label{PLduality2}
The following isomorphism is valid:   
\[
H^2({\cal N}_-\cup{\cal N}_+,{\cal N}_+;\ze_2)
\cong H_{d-2}({\cal N}_-^\ast,\partial{\cal N}_-^\ast
\cap\partial\Lambda^\ast;\ze_2),
\]
where the right-hand side is the set of all the homology classes of 
the $(d-2)$-complexes $\Gamma$ such that $\Gamma\cap{\cal N}_+=\emptyset$ 
and that $\partial\Gamma\subset\partial{\cal N}_-^\ast
\cap\partial\Lambda^\ast$.  
\end{prop}

\begin{proof}
Let $\eta$ be a two-cochain. Then we can define 
the $(d-2)$-dimensional complex $\Gamma$ which consists of the $(d-2)$-cells 
which is dual to the plaquettes $p$ having the plaquette variable $\eta_p=-1$. 
This is an extension of the dual map from frustration networks to  
$(d-2)$-dimensional complexes. 
We write $\Gamma=\zeta(\eta)$ for this map $\eta\longrightarrow \Gamma$. 
This map $\zeta$ gives the desired isomorphism of 
the proposition. 
Let $\eta,\eta'\in Z^2({\cal N}_-\cup{\cal N}_+,{\cal N}_+;\ze_2)$ satisfy 
$\eta'\sim\eta$. First we want to show $\zeta(\eta')\sim\zeta(\eta)$. 
Namely the map of the proposition is well defined. From the definitions of 
the two-cocycles, there exists bond variables $\epsilon=\{\epsilon_{ij}\}$ satisfying  
\[
\eta_p'=\eta_p\prod_{\langle i,j\rangle\subset p}\epsilon_{ij}
\]
with the condition $\epsilon_{ij}=1$ for $\langle i,j\rangle\subset{\cal N}_+$. 
Therefore one has $\zeta(\eta')=\zeta(\eta)+\zeta(\epsilon)$. 
The collection of the $(d-1)$-cells which are dual to the bonds $\langle i,j\rangle$ 
with $\epsilon_{ij}=-1$ forms $(d-1)$-dimensional hypersurfaces $\Sigma$. 
Since $\Sigma$ ends at a plaquette $p$ satisfying $\epsilon(\partial p)=-1$ 
or at $\partial{\cal N}_-^\ast\cap\partial\Lambda^\ast$, 
the boundary $\partial\Sigma$ is made of the collection of the $(d-2)$-cells 
which are dual to the plaquettes $p$ satisfying $\epsilon(\partial p)=-1$ 
or end at $\partial{\cal N}_-^\ast\cap\partial\Lambda^\ast$. 
Namely $\partial\Sigma=\zeta(\epsilon)+\Gamma_0$ 
with $\Gamma_0\subset\partial{\cal N}_-^\ast\cap\partial\Lambda^\ast$. 
This implies $\zeta(\epsilon)\sim 0$. 
Immediately, $\zeta(\eta')\sim\zeta(\eta)$. 

Let $\Gamma\subset{\cal N}_-^\ast$ be a $(d-2)$-dimensional complex satisfying 
$\partial\Gamma\subset\partial{\cal N}_-^\ast\cap\partial\Lambda^\ast$. 
If a cube $c\subset{\cal N}_-\cup{\cal N}_+$ intersects $\Gamma$, 
an even number of the plaquettes of $c$ must intersect $\Gamma$. 
For such a plaquette $p$, we choose the plaquette variables $\eta_p=-1$. 
Then the corresponding two-cochain $\eta$ satisfies the cocycle 
condition $\eta(\partial c)=1$ for any cube $c$, 
and $\zeta(\eta)=\Gamma$. Thus the map of the proposition is surjective. 

Finally we show that the map is also injective. 
Let $\eta\in Z^2({\cal N}_-\cup{\cal N}_+,{\cal N}_+;\ze_2)$ satisfy $\zeta(\eta)\sim 0$. 
Then there exists a $(d-1)$-dimensional hypersurface $\Sigma\subset{\cal N}_-^\ast$ 
such that $\partial\Sigma=\zeta(\eta)+\Gamma_0$ 
with $\Gamma_0\subset\partial{\cal N}_-^\ast\cap\partial\Lambda^\ast$. 
We take bond variables $\epsilon=\{\epsilon_{ij}\}$ so that 
$\epsilon_{ij}=-1$ if the bond $\langle i,j\rangle$ pierces $\Sigma$, and  
$\epsilon_{ij}=+1$ otherwise. It is sufficient to show 
\[
\eta_p\prod_{\langle i,j\rangle\subset p}\epsilon_{ij}=1.
\]
When $\eta_p=-1$, an odd number of the bond of $p$ pierces $\Sigma$ because 
$\partial\Sigma$ ends at $p$. When $\eta_p=+1$, an even number of the bond of $p$ 
pierces $\Sigma$. Thus, in both of the two cases, the above relation holds.  
\end{proof}

The main results of this section is summarized as:

\begin{thm}
\label{thm:main}
We have the commutative diagram: 
\[
\CD
H^1({\cal N}_+;\ze_2) @>\partial^\ast>> H^2({\cal N}_-\cup{\cal N}_+,{\cal N}_+;\ze_2)\\
@VV \cong V @VV \cong V\\
H_{d-1}({\cal N}_+^\ast,\partial{\cal N}_+^\ast;\ze_2) @>\partial >> 
H_{d-2}({\cal N}_-^\ast,\partial{\cal N}_-^\ast\cap\partial\Lambda^\ast;\ze_2)
\endCD
\]
\end{thm}

\begin{proof}
Let $\phi\in Z^1({\cal N}_+;\ze_2)$ with the bond variables ${\hat J}_{ij}$, 
and let $\{\sigma_i\}_{i\in{\cal N}_+}$ be a spin configuration of the ground state of 
the Hamiltonian on ${\cal N}_+\cup{\cal N}_-$. 
Then the cocycle $\phi'=\{{\hat J}_{ij}'={\hat J}_{ij}\sigma_i\sigma_j\}$ determines 
the domain walls as we showed in Proposition~\ref{PLduality1}, 
and the boundaries of the domain walls are the $(d-2)$-dimensional complexes. 

On the other hand, since $\partial^\ast\phi'$ is written in terms of only 
the bond variables ${\hat J}_{ij}'$, we can find a two-cocycle $\eta$ 
which is cohomologous to $\partial^\ast\phi'$ and written in terms of 
only the bond variables $\tau_{ij}$ 
which satisfy $\tau_{ij}={\hat J}_{ij}'$ for $\langle i,j\rangle\subset{\cal N}_+$, 
and $\tau_{ij}=1$ otherwise. Therefore the bond variables $\tau_{ij}={\hat J}_{ij}'=-1$ on 
$\partial{\cal N}_+$ yield the frustrated plaquettes $p$ with 
$\eta_p=-1$ in ${\cal N}_-$. These frustrated plaquettes determine the same 
$(d-2)$-dimensional complexes as the above complexes by Proposition~\ref{PLduality2}.  
\end{proof}

\section{Homology of domain walls}
\label{homoDW}

Relying on Theorem~\ref{thm:main}, we discuss the topology of the domain walls 
for a ground state. 

Consider a generic spin configuration $\{\sigma_i\}$ 
on $\Lambda={\cal N}_+\cup{\cal N}_-$. 
We write ${\hat E}_{ij}={\hat J}_{ij}\sigma_i\sigma_j$. 
Then the sign of the bond energy is given by $-{\hat E}_{ij}$. 
Consider the collection of all the $(d-1)$-cells which are dual to the bonds 
$\langle i,j\rangle$ having ${\hat E}_{ij}=-1$. 
Then the domain walls for the spin configuration 
are given by the connected elements of the collection.   
If a plaquette $p$ is frustrated, then the number of the bonds having 
${\hat E}_{ij}=-1$ in $p$ is odd. 
Therefore the corresponding domain wall ends at the plaquette $p$. 
The $(d-2)$-cell which is dual to the plaquette $p$ 
becomes the boundary of the domain wall. If $p$ is unfrustrated, 
the number of the bonds having ${\hat E}_{ij}=-1$ in $p$ becomes even.  
In this case, the boundary of the domain wall does not appear at the plaquette $p$. 
Thus a domain wall is a connected, $(d-1)$-dimensional hypersurface 
whose boundary is a collection of $(d-2)$-dimensional complexes 
dual to the frustrated plaquettes 
or ends at the boundary of the lattice $\Lambda^\ast$. 

{From} the proof of Theorem~\ref{thm:main}, one notices the following fact: 
All of the domain walls are outside the unfrustration network  
except for the domain walls which are transverse to a frustrated loop 
in the unfrustration network. In order to show this, we introduce 
the bond variables $\tau_{ij}'$ which satisfy 
$\tau_{ij}'=1$ for $\langle i,j\rangle\subset{\cal N}_+$, 
and $\tau_{ij}'={\hat J}_{ij}'$ otherwise. 
Clearly one has the decomposition, ${\hat J}_{ij}'={\hat J}_{ij}\sigma_i\sigma_j=
\tau_{ij}\tau_{ij}'$, 
where $\tau_{ij}$ is the cochain in the proof of Theorem~\ref{thm:main}. This relation implies that 
$\tau_{ij}$  yields the set of the domain walls $\theta\circ\kappa(\tau)$ 
which are transverse to a frustrated loop in the unfrustration network ${\cal N}_+$ 
as we have seen in the proof of Theorem~\ref{thm:main}, and that 
$\tau_{ij}'$ yields the rest of the domain walls $\theta\circ\kappa(\tau')$ 
which are outside the unfrustration network ${\cal N}_+$ 
from the condition $\tau_{ij}'=1$ for $\langle i,j\rangle\subset{\cal N}_+$. 

To summarize, we obtain the following description of the domain walls for the ground state. 
The homology class of the hypersurfaces $\theta\circ\kappa(\tau)$ 
is determined by the frustrated loops on ${\cal N}_+$ as in the relation (\ref{H1Hd-1}). 
Besides, the size of the hypersurfaces $\theta\circ\kappa(\tau)$ cannot become large 
for the ground state because the position and profile of the domain walls are determined 
to minimize the total domain wall energy. 
On the other hand, the rest of the domain walls $\theta\circ\kappa(\tau')$ 
are not expected to show such a similar, simple structure 
as we showed in Section~\ref{topofru}. 
But the outstanding feature is that they are all outside the unfrustration network. 

The spins on the unfrustration network prefer to maintain 
their relative orientation 
although the neighboring pairs of the spins for the bonds on the domain walls 
cannot take their favorable orientation. Therefore 
we can expect that the ground state exhibits order of the frozen spins 
on the unfrustration network even if the size of the frustration network is large. 
On the frustration network, there appear many domain walls whose boundaries end 
at a frustrated plaquette. 
If the ground state on the frustration network is highly degenerate, 
then the spin configurations yield many patterns of the domain walls 
on the frustration network. 
In such a situation, order of the frozen spins on the frustration network cannot be expected. 
 
\section{Topological effect and the spin glass phase}
\label{Perco}

In this section, we discuss the role of topologically 
nontrivial domain walls which are transverse to a frustrated loop in 
the unfrustration network in the context of the appearance of 
the spin glass phase at finite temperatures.  

\subsection{Absence of the spin glass phase in two dimensions}

Consider first the system on the square lattice $\ze^2$. 
As we will see below, a naive application of a percolation argument as in \cite{BF} 
to the unfrustration network yields the existence of long-range order of the frozen spins 
on the unfrustration network when ignored the effect of the domain walls 
which are transverse to a frustrated loop on the unfrustration network.
However, it is widely believed that there is no spin glass phase in two dimensions 
at finite temperatures. Thus the percolation argument alone cannot explain the absence of 
the spin glass phase in two dimensions. 
We expect that the thermal fluctuations of the domain walls which are 
transverse to a frustrated loop on the unfrustration network, plays an essential role 
in the growth process of the long-range order. 
Namely, the thermal fluctuations of the topologically nontrivial 
domain walls can be expected 
to destroy the long-range order of the frozen spins on the unfrustration network 
in two dimensions.  
This is essentially global, topological effect for the unfrustration network because 
the sizes of the domain walls are expected to be ignorably 
small in the unfrustration network. 

To begin with, let us estimate the cluster sizes of the unfrustration network 
by using percolation argument \cite{BF}. 
Let $A_n$ be a set of $n$ plaquettes. We denote by $\alpha_n$ the event 
that all the plaquettes of $A_n$ are unfrustrated. By definition, 
one can decompose any set $A_n$ into a plaquette $p$ and a set $A_{n-1}$ 
such that $p$ and $A_{n-1}$ share at most two bonds. 
Write $J_1,J_2,J_3,J_4$ for the four bonds of $\partial p$, and 
assume that the two bonds, $J_3,J_4$, are not included in the complex 
which is made of $A_{n-1}$. 
Then the probability that all the plaquettes of $A_n$ are unfrustrated 
is written 
\[
{\rm Prob}[\alpha_n]=\int\prod_{\langle i,j\rangle\subset(A_{n-1}\backslash\partial p)}
d\rho(J_{ij})\int\prod_{k=1}^4 d\rho(J_k)\ \chi[\alpha_{n-1}]\ \chi[\phi(\partial p)=+1],
\]
where $\chi[\alpha]$ is the indicator function of an event $\alpha$. 
In the same way as in \cite{BF}, one has 
\begin{align*}
{\rm Prob}[\alpha_n]&=\int\prod_{\langle i,j\rangle\subset(A_{n-1}\backslash\partial p)}
d\rho(J_{ij})\int dJ_1dJ_2\; \chi[\alpha_{n-1}]\\
&\times\left\{[g(-J_1)g(-J_2)(1-x)^2+g(J_1)g(J_2)x^2]\cdot[x^2+(1-x)^2]\right.\\
&+\left.[g(-J_1)g(J_2)+g(J_1)g(-J_2)]\cdot x(1-x)\cdot2x(1-x)\right\}\\
&={\rm Prob}[\alpha_{n-1}]\cdot 2x(1-x)\\
&+\int\prod_{\langle i,j\rangle\subset(A_{n-1}\backslash\partial p)}
d\rho(J_{ij})\int dJ_1dJ_2\; \chi[\alpha_{n-1}]\\
&\times[g(-J_1)g(-J_2)(1-x)^2+g(J_1)g(J_2)x^2](2x-1)^2.
\end{align*}
Since the second term is positive and vanishes for $x=1/2$, one obtains 
\[
{\rm Prob}[\alpha_n]\ge [2x(1-x)]^n
\] 
and 
\[
{\rm Prob}[\alpha_n]=(1/2)^n \quad\mbox{for }\ x=1/2.
\]
{From} these results, one obtains the following:   
For $x$ near $1/2$, there exists an infinite, connected cluster of unfrustrated 
plaquettes. Here two plaquettes are considered connected if they have at least 
one point in common. 
This result by percolation argument suggests the existence of long range order 
of the frozen spins on the unfrustration network. 

To examine this expectation, fix $x=1/2$ for simplicity, and 
consider first the Hamiltonian whose interactions are restricted onto 
the bonds of the unfrustration network. 
To begin with, we recall the following fact 
about dilute Ising ferromagnets \cite{ACCN,ACCN2}: 
If all of the couplings $J_{ij}$ are equal to $J_0>0$, 
then the system shows ferromagnetic long-range order. 
As we showed in Theorem~\ref{thm:unfrusGS}, all the bonds except for 
the bonds which intersect a domain wall have a favorable energy 
in a ground-state configuration. 
Here all of the domain walls must be transverse to 
a frustrated loop on the unfrustration network. 
Besides, the position and profile of the domain walls 
for the ground state are determined to minimize the total size of the domain walls. 
{From} these observations, the total size of the domain walls is expected to be ignorably 
small on the unfrustration network. 
Namely the bonds having unfavorable energies are expected 
to be a small fraction of the bonds of the unfrustration network. 
If we can ignore the presence of the domain walls on the unfrustration network, 
the Hamiltonian restricted onto the unfrustration network is equivalent to 
that of the standard ferromagnetic Ising model on the network 
by a gauge transformation. 

But the fluctuations of the domain walls which arise from the frustration networks 
may destroy the long-range order on the unfrustration network as a boundary effect. 
Clearly the domain walls on the frustration network also minimize their sizes 
in a ground state. This implies that the domain walls show a tendency to 
confine themselves into a small neighborhood of the frustration network.  
In other words, the long-range order on the unfrustration network 
is expected to be stable against the fluctuations of the domain walls on 
the frustration network.  
Thus the percolation argument naively leads to 
the existence of long-range order on the unfrustration network 
when ignored the effect of the domain walls 
which are transverse to a frustrated loop on the unfrustration network. 
In other words, the percolation argument alone cannot explain 
the absence of the spin glass phase in two dimensions at finite temperatures. 

Now we take account of the effect of the thermal fluctuations 
of the topologically nontrivial domain walls on the unfrustration network. 
For this purpose, we recall the well known fact that a large domain wall is not stable 
in the Ising ferromagnet on the $\ze^2$ lattice at finite temperatures \cite{Gallavotti}. 
Consider a large cluster of unfrustrated plaquettes.  
If there appears a large domain wall which is transverse to 
a large frustrated loop on the cluster, then one can expect that 
the thermal fluctuation of the domain wall destroys the long-range order 
of the cluster. Besides, we cannot expect that the probability that such a domain wall 
appear on a cluster of unfrustrated plaquettes, is negligibly small. 
Actually, if the probability is negligibly small, then 
the frustrated loop which is transverse to the domain wall must go trough 
the inside of a small handle with probability nearly equal to one. 
But one can expect that there appear many large handles for a sufficiently large cluster 
of unfrustrated plaquettes with nonnegligible probability. 
In consequence, we can expect that 
the thermal fluctuation destroys the long range order at finite temperatures. 
 
\subsection{The spin glass phase in three and higher dimensions}

We specialize to the simple cubic lattice $\ze^3$. Our argument in 
higher dimensions or on other lattices is the same. In the same way as in 
Bovier and Fr\"ohlich \cite{BF}, one can show that 
the unfrustration network percolates for $x$ near 1/2. 
When we restrict the present system to the unfrustration network 
with infinite volume, the situation is very similar to that of 
the Ising ferromagnet except that there appear many topologically nontrivial domain walls 
on the unfrustration network. 
But, unlike the square lattice $\ze^2$, 
the domain walls in the Ising ferromagnet is 
known to be stable against the thermal fluctuation on $\ze^3$ at low temperatures 
\cite{Dobrushin}. Thus one can expect that such topologically nontrivial domain walls 
are stable against the thermal fluctuation on $\ze^3$ at low temperatures, too. 
These observations suggest that 
the thermal fluctuation of the topologically nontrivial domain walls 
cannot destroy the long-range order of the frozen spins on the unfrustration network. 

There remains the possibility that the long-range order is 
ferromagnetic or antiferromagnetic order except for the case with $x=1/2$. 
Let us preclude the possibility.  For simplicity, we assume $x>1/2$ but near 1/2.  
As is well known, the critical density of a Bernoulli bond percolation process 
on the cubic lattice \cite{Stauffer} is strictly less than 1/2. 
Therefore the bonds of negative couplings $J_{ij}$ percolate. 
Combining this with the discussion after Lemma~\ref{lem:int}, we can expect that 
there appear many interfaces between up and down spins on the unfrustration network 
with infinite volume. 
This precludes the possibility of the ferromagnetic order on the unfrustration network. 
Thus we can expect the existence of 
the spin glass phase on the $\ze^3$ lattice for the density $x$ near 1/2.  

\appendix

\section{Proof of the relation (\ref{frustratedloop})}
\label{UCT}

In order to make the paper self-contained, we give a proof of 
(\ref{frustratedloop}). 

For a map $f$, we denote by ${\rm Im}\;f$ the image of the map $f$, and 
denote by ${\rm Ker}\;f$ the kernel of $f$. 
Let $\Lambda\subset\ze^d$ be a collection of plaquettes. 
Consider the exact sequence on $\Lambda$:
\begin{equation}
0\longrightarrow B_1(\Lambda;\ze)\mathop{\longrightarrow}^i Z_1(\Lambda;\ze)
\mathop{\longrightarrow}^p H_1(\Lambda;\ze)\longrightarrow 0
\label{esBZH}
\end{equation}
Namely, ${\rm Ker}\;i=0$, ${\rm Im}\;i={\rm Ker}\;p$ and the map $p$ is surjective.  
Here the map $i$ is the inclusion, $B_1(\Lambda;\ze)\subset Z_1(\Lambda;\ze)$, 
and $p$ is the projection.
We write 
\[
H_1^\#={\rm Hom}(H_1(\Lambda;\ze),\ze_2),
\]
\[
Z_1^\#={\rm Hom}(Z_1(\Lambda;\ze),\ze_2),
\]
and
\[
B_1^\#={\rm Hom}(B_1(\Lambda;\ze),\ze_2).
\]
We choose the additive group $\ze_2=\{0,1\}$ for the representation of $\ze_2$. 
We define the map $p^\#:H_1^\#\longrightarrow Z_1^\#$ by the adjoint $p^\#$ of 
the projection $p$ as 
\[
(p^\#\alpha)(\ell)=\alpha(p(\ell))
\]
for $\alpha\in H_1^\#$ and $\ell\in Z_1(\Lambda;\ze)$. Similarly we define 
$i^\#:Z_1^\#\longrightarrow B_1^\#$ by 
\[
(i^\# z)(\partial s)=z(i(\partial s))
\]
for $z\in Z_1^\#$ and $\partial s\in B_1(\Lambda;\ze)$. 

\begin{lem}
The following sequence is exact:
\[
0\longrightarrow H_1^\#\mathop{\longrightarrow}^{p^\#}Z_1^\#
\mathop{\longrightarrow}^{i^\#} B_1^\#
\]
\end{lem}

\begin{proof}
First we show ${\rm Ker}\;p^\#=0$. Assume $p^\#\alpha=0$ for $\alpha\in H_1^\#$. 
{From} the exact sequence (\ref{esBZH}), we have that, for any $a\in H_1(\Lambda;\ze)$, 
there exists $\ell\in Z_1(\Lambda;\ze)$ such that $a=p(\ell)$. Then one obtains 
\[
0=(p^\#\alpha)(\ell)=\alpha(p(\ell))=\alpha(a) 
\]
for any $a\in H_1(\Lambda;\ze)$. This implies $\alpha=0$. 

Next we show ${\rm Im}\;p^\#\subset {\rm Ker}\; i^\#$. Let $\alpha\in H_1^\#$ and 
$\partial s\in B_1(\Lambda;\ze)$. Then one has 
\[
(i^\#\circ p^\#\alpha)(\partial s)=(p^\#\alpha)(i(\partial s))=
\alpha(p\circ i(\partial s))=0
\]
from the exact sequence (\ref{esBZH}). 

Finally we show ${\rm Ker}\; i^\#\subset{\rm Im}\;p^\#$. Assume $i^\# z=0$ 
for $z\in Z_1^\#$. Let $\ell,\ell'\in Z_1(\Lambda;\ze)$ such that $p(\ell)=p(\ell')$. 
Then $\ell-\ell'\in {\rm Ker}\; p={\rm Im}\; i$ from (\ref{esBZH}). 
Combining this with the assumption $i^\# z=0$, one has $z(\ell-\ell')=0$. 
Thus, if $p(\ell)=p(\ell')$, then $z(\ell)=z(\ell')$. 
This defines $\alpha\in H_1^\#$ by 
\[
\alpha(p(\ell))=z(\ell). 
\]
This left-hand side is equal to $(p^\#\alpha)(\ell)$. 
\end{proof}

Clearly this lemma yields  the following isomorphism: 
\begin{equation}
H_1^\#={\rm Hom}(H_1(\Lambda;\ze),\ze_2)\cong {\rm Ker}\;i^\#
\label{H1congKeri} 
\end{equation}

We denote by $B_0(\Lambda;\ze)$ the module made of the boundaries $\partial\gamma$ 
for all the oriented path $\gamma$ in $\Lambda$ with the coefficients $\ze$. 
Consider the exact sequence, 
\begin{equation}
0\longrightarrow Z_1(\Lambda;\ze)\mathop{\longrightarrow}^j
C_1(\Lambda;\ze)\mathop{\longrightarrow}^\partial B_0(\Lambda;\ze)
\longrightarrow 0
\label{esZCB}
\end{equation}
where the map $j$ is the inclusion $Z_1(\Lambda;\ze)\subset C_1(\Lambda;\ze)$. 
We write 
\[
B_0^\#={\rm Hom}(B_0(\Lambda;\ze),\ze_2)
\]
and
\[
C_1^\#={\rm Hom}(C_1(\Lambda;\ze),\ze_2).
\]
We define the map $\partial^\#:B_0^\#\longrightarrow C_1^\#$ by 
the adjoint $\partial^\#$ of the boundary operator $\partial$ as 
\[
(\partial^\# \beta)(c)=\beta(\partial c)
\]
for $\beta\in B_0^\#$ and $c\in C_1(\Lambda;\ze)$. Further we define 
$j^\#:C_1^\#\longrightarrow Z_1^\#$ by 
\[
(j^\#f)(\ell)=f(j(\ell))
\]
for $f\in C_1^\#$ and $\ell\in Z_1(\Lambda;\ze)$. 

\begin{lem}
\label{lemma:esBCZ}
The following exact sequence is valid:
\[
0\longrightarrow B_0^\# \mathop{\longrightarrow}^{\partial^\#}C_1^\#
\mathop{\longrightarrow}^{j^\#}Z_1^\#\longrightarrow 0
\]
\end{lem}

\begin{proof}
First we show ${\rm Ker}\;\partial^\#=0$. Assume $\partial^\#\beta=0$ for $\beta\in B_0^\#$. 
{From} the exact sequence (\ref{esZCB}), one has that for any $b\in B_0(\Lambda;\ze)$, 
there exists $c\in C_1(\Lambda;\ze)$ such that $b=\partial c$.  Therefore one obtains 
\[
0=(\partial^\#\beta)(c)=\beta(\partial c)=\beta(b)
\]
for any $b\in B_0(\Lambda;\ze)$. This implies $\beta=0$. 

Next we show ${\rm Im}\; \partial^\#={\rm Ker}\;j^\#$. 
Let $\beta\in B_0^\#$ and $\ell\in Z_1(\Lambda;\ze)$. Then 
\[
(j^\#\circ \partial^\#\beta)(\ell)=\beta(\partial\circ j(\ell))=0.
\]
This implies ${\rm Im}\;\partial^\#\subset{\rm Ker}\;j^\#$.  

In order to prove ${\rm Ker}\; j^\#\subset{\rm Im}\;\partial^\#$, we assume 
$j^\#f=0$ for $f\in C_1^\#$. Let $c,c'\in C_1(\Lambda;\ze)$ such that 
$\partial c=\partial c'$. Then one has $c-c'\in {\rm Ker}\;\partial={\rm Im}\;j$ 
{from} the exact sequence (\ref{esZCB}). 
Combining this with the assumption $j^\#f=0$, 
one has $f(c-c')=0$. Thus, if $\partial c=\partial c'$, then $f(c)=f(c')$. 
This defines $\beta\in B_0^\#$ by 
\[
\beta(\partial c)=f(c). 
\]
This left-hand side is equal to $(\partial^\#\beta)(c)$. 
Therefore ${\rm Ker}\;j^\#\subset {\rm Im}\;\partial^\#$. 

Combining this with the above ${\rm Im}\;\partial^\#\subset{\rm Ker}\;j^\#$, 
the desired result ${\rm Im}\;\partial^\#={\rm Ker}\;j^\#$ is obtained. 

Finally we show that the map $j^\#$ is surjective. 
Let $\{b_\lambda\}_\lambda$ is a basis of the module $B_0(\Lambda;\ze)$. 
Then, for each $\b_\lambda$, there exists a chain $c_\lambda\in C_1(\Lambda;\ze)$ such that 
$b_\lambda=\partial c_\lambda$. This defines the map 
\[
\overline{\partial}:B_0(\Lambda;\ze)\longrightarrow C_1(\Lambda;\ze)
\]
One can notice that any chain $c\in C_1(\Lambda;\ze)$ can be decomposed into two parts as 
$c=\overline{\partial}(b)+j(\ell)$ 
with $b\in B_0(\Lambda;\ze)$ and $\ell\in Z_1(\Lambda;\ze)$. Using this decomposition, 
we define the map $\overline{j}:C_1(\Lambda;\ze)\longrightarrow Z_1(\Lambda;\ze)$ 
by $\overline{j}(c)=\ell$. Clearly one has $\overline{j}\circ j=1$.   
Further we can define the map $\overline{j}^\#:Z_1^\#\longrightarrow C_1^\#$ by 
\[
(\overline{j}^\# z)(c)=z(\overline{j}(c))
\]
for $z\in Z_1^\#$ and $c\in C_1(\Lambda;\ze)$. Then one has $j^\#\circ\overline{j}^\#=1$. 
Actually one can easily show 
\[
(j^\#\circ\overline{j}^\#z)(\ell)=(\overline{j}^\#z)(j(\ell))=z(\overline{j}\circ j(\ell))
=z(\ell) 
\]
for any $z\in Z_1^\#$ and for any loop $\ell\in Z_1(\Lambda;\ze)$. 
The result, $j^\#\circ\overline{j}^\#=1$, implies that the map $j^\#$ is surjective.
\end{proof}

\begin{lem}
The following sequence is exact:
\[
0\longrightarrow H^1(\Lambda;\ze_2)\mathop{\longrightarrow}^{j^\ast}
Z_1^\#\mathop{\longrightarrow}^{i^\#}B_1^\# 
\]
\end{lem}

\begin{proof}
We choose the multiplicative group $\ze_2=\{1,-1\}$ for the representation of $\ze_2$. 
Let $\alpha\in B^1(\Lambda;\ze_2)$. Then $(j^\ast\alpha)(\ell)=\alpha(j(\ell))=1$ 
for any $\ell\in Z_1(\Lambda;\ze)$. 
Thus the adjoint $j^\ast$ of 
the inclusion $j:Z_1(\Lambda;\ze)\subset C_1(\Lambda;\ze)$ is well defined. 

Let us show ${\rm Ker}\; j^\ast=0$. Assume $j^\ast\alpha=1$ for $\alpha\in H^1(\Lambda;\ze_2)$. 
Then one has 
\[
1=(j^\ast\alpha)(\ell)=\alpha(j(\ell))=\alpha(\ell)
\]
for any loop $\ell\in Z_1(\Lambda;\ze)$. Therefore we have $\alpha\in B^1(\Lambda;\ze_2)$ 
{from} the proof of Lemma~\ref{lemma:nofrustration}. 

Next we show ${\rm Im}\; j^\ast\subset {\rm Ker}\; i^\#$. Let $\alpha\in H^1(\Lambda;\ze_2)$ 
and $\partial s\in B_1(\Lambda;\ze)$. Then 
\[
(i^\#\circ j^\ast\alpha)(\partial s)=(j^\ast\alpha)(i(\partial s))
=\alpha(j\circ i(\partial s))=\alpha(\partial s)=(\partial^\ast\alpha)(s)=1. 
\]

Finally we show ${\rm Ker}\;i^\#\subset{\rm Im}\; j^\ast$. 
Assume $i^\#z=1$ for $z\in Z_1^\#$. Then one has 
\[
1=(i^\#z)(\partial s)=z(i(\partial s))=z(\partial s).
\]
{From} Lemma~\ref{lemma:esBCZ}, there exists $f\in C_1^\#$ such that $z=j^\#f$. 
Substituting this into the above equality, one obtains 
\[
1=(j^\#f)(\partial s)=f(j(\partial s))=f(\partial s)=(\partial^\ast f)(s)
\]
for any $s$. 
This implies that $f$ can be identified with $\alpha\in H^1(\Lambda;\ze_2)$. 
Thus $z=j^\#\alpha$. 
\end{proof}

Clearly this lemma yields the isomorphism:
\[
H^1(\Lambda;\ze_2)\cong{\rm Im}\;j^\ast ={\rm Ker}\; i^\#
\]
Combining this with the isomorphism (\ref{H1congKeri}), one obtains the desired isomorphism: 
\[
H^1(\Lambda;\ze_2)\cong {\rm Hom}(H_1(\Lambda;\ze);\ze_2)
\]




\begin{thebibliography}{99}

\bibitem{ACCN} Aizenman, M., Chayes, J. T., Chayes, L. and Newman, C. M.: 
The phase boundary in dilute and random Ising and Potts ferromagnets. 
J. Phys. A: Math. Gen. {\bf 20} (1987), L313--L318. 

\bibitem{ACCN2} Aizenman, M., Chayes, J. T., Chayes, L. and Newman, C. M.: 
Discontinuity of the magnetization in one-dimensional $1/|x-y|^2$ Ising 
and Potts models, 
J. Stat. Phys. {\bf 50} (1988), 1--40.

\bibitem{ARS} Avron, J. E., Roepstorff, G. and Schulman, L. S.: 
Ground state degeneracy and ferromagnetism in a spin glass. 
J. Stat. Phys. {\bf 26} (1981), 25--36.

\bibitem{BF} Bovier, A. and Fr\"ohlich, J.:
A heuristic theory of the spin glass phase.
J. Stat. Phys. {\bf 44} (1986), 347--391.

\bibitem{Dobrushin} Dobrushin, R. L.: Gibbs states describing the coexistence of 
the phases for a three-dimensional Ising model. 
Theor. Prob. Appl. {\bf 17} (1972), 582--600. 

\bibitem{DFN} Dubrovin, B. A., Fomenko, A. T. and Novikov, S. P.: 
Modern Geometry--- Methods and Applications, Part ~III. Introduction to 
Homology Theory. 
Springer-Verlag, New York, 1990.

\bibitem{EA} Edwards, S. and Anderson, P. W.:
Theory of spin glasses. 
J. Phys. F{\bf 5} (1975), 965--974.

\bibitem{vEMN} van Enter, A. C. D., Medved', I. and Neto${\check {\rm c}}$n\'y, K.: 
Chaotic size dependence in the Ising model with random boundary conditions. 
Markov Proc. Rel. Fields. {\bf 8} (2002) 479--508.

\bibitem{vENS} van Enter, A. C. D., Neto${\check {\rm c}}$n\'y, K. and Schaap, H. G.: 
On the Ising model with random boundary condition. 
J. Stat. Phys. {\bf 118} (2005) 997--1057. 

\bibitem{ES} Eilenberg, S. and Steenrod, N.: Foundations of Algebraic Topology. 
Princeton Univ. Press, Princeton, New Jersey, 1952. 

\bibitem{Gallavotti} Gallavotti, G.: The phase separation line 
in the two-dimensional Ising model. Commun. Math. Phys. {\bf 27} (1972) 103--136. 

\bibitem{Greenberg} Greenberg, M.: Lectures on Algebraic Topology. 
W. A. Benjamin, New York, 1966.

%
%

\bibitem{MPV} Mezard, M., Parisi, G. and Virasoro, M. A.: 
Spin glass theory and beyond. World Scientific, Singapore, 1987.  

\bibitem{Panati} Panati, G.: Triviality of Bloch and Bloch-Dirac bundles. 
Ann. Henri Poincar\'e \ {\bf 8} (2007), 995--1011. 

\bibitem{Spanier} Spanier, E. H.: Algebraic Topology. 
Springer-Verlag, New York, 1966. 

\bibitem{Stauffer} Stauffer, D.: Introduction to percolation theory. 
Taylor \& Francis, London and Philadelphia, 1985. 

\bibitem{SBLB} Stein, D. L., Baskaran, G, Liang, S. and Barber, M. N.: 
Ground state of the $\pm J$ Ising spin glass. Phys. Rev. B{\bf 36} (1987), 5567--5571.  

\bibitem{Toulouse} Toulouse, G.:
Theory of the frustration effect in spin glasses: I. 
Commun. Phys. {\bf 2} (1977), 115--119.

\end{thebibliography}
\end{document}